\documentclass[english,aps,prb,amsmath,amsfonts,superscriptaddress, twocolumn]{revtex4-1}
\usepackage[T1]{fontenc}
\usepackage[latin9]{inputenc}
\usepackage{array}
\usepackage{units}
\usepackage{textcomp}
\usepackage{multirow}
\usepackage{amsmath}
\usepackage{amssymb}
\usepackage{graphicx}
\usepackage{subscript}

\makeatletter

\providecommand{\tabularnewline}{\\}

\usepackage{marginnote}

\makeatother

\usepackage{babel}
\begin{document}
\title{Unconventional photo-induced charge-density-wave dynamics in 2$H$-NbSe$_{2}$}
\author{R. Venturini}
\affiliation{Department of Complex Matter, Jozef Stefan Institute, Jamova 39, 1000
Ljubljana, Slovenia}
\author{A. Sarkar}
\affiliation{Department of Complex Matter, Jozef Stefan Institute, Jamova 39, 1000
Ljubljana, Slovenia}
\author{P. Sutar}
\affiliation{Department of Complex Matter, Jozef Stefan Institute, Jamova 39, 1000
Ljubljana, Slovenia}
\author{Z. Jagli\v{c}i\'{c}}
\affiliation{Faculty of Civil and Geodetic Engineering, University of Ljubljana,
Jamova cesta 2, Ljubljana, Slovenia}
\affiliation{Institute of Mathematics, Physics and Mechanics, Jadranska 19, Ljubljana,
Slovenia}
\author{Y. Vaskivskyi}
\affiliation{Department of Complex Matter, Jozef Stefan Institute, Jamova 39, 1000
Ljubljana, Slovenia}
\author{E. Goreshnik}
\affiliation{Dept. of Inorganic Chemistry and Technology, Jozef Stefan Institute,
Jamova 39, 1000 Ljubljana, Slovenia}
\author{D. Mihailovic}
\affiliation{Department of Complex Matter, Jozef Stefan Institute, Jamova 39, 1000
Ljubljana, Slovenia}
\affiliation{Center of Excellence on Nanoscience and Nanotechnology Nanocenter
(CENN Nanocenter), Jamova 39, 1000 Ljubljana, Slovenia}
\author{T. Mertelj}
\email{tomaz.mertelj@ijs.si}

\affiliation{Department of Complex Matter, Jozef Stefan Institute, Jamova 39, 1000
Ljubljana, Slovenia}
\affiliation{Center of Excellence on Nanoscience and Nanotechnology Nanocenter
(CENN Nanocenter), Jamova 39, 1000 Ljubljana, Slovenia}
\date{\today}
\begin{abstract}
We investigated temperature ($T$) dependent ultrafast near-infrared
(NIR) transient reflectivity dynamics in coexisting superconducting
(SC) and charge density wave (CDW) phases of two-dimensional 2\emph{H}-NbSe$_{2}$
using NIR and visible excitations. With visible pump-photon excitation
(400 nm) we find a slow high-energy quasiparticle relaxation channel
which is present in all phases. In the CDW phase, we observe a distinctive
transient response component, irrespective of the pump-photon energy.
The component is marked by the absence of coherent amplitude mode
oscillations and a relatively slow, picosecond rise time, which is
different than in most of the typical CDW materials. In the SC phase,
another tiny component emerges that is associated with optical suppression
of the SC phase.

The transient reflectivity relaxation in the CDW phase is dominated
by phonon diffusive processes with an estimated low-$T$ heat diffusion
constant anisotropy of $\sim30$.

 Strong excitation of the CDW phase reveals a weakly non-thermal CDW
order parameter (OP) suppression. Unlike CDW systems with a larger
gap, where the optical OP suppression involves only a small fraction
of phonon degrees of freedom, the OP suppression in 2\emph{H}-NbSe$_{2}$
is characterised by the excitation of a large amount of phonon degrees
of freedom and significantly slower dynamics. 
\end{abstract}
\maketitle

\section{introduction}

Layered materials with reduced dimensionality offer a platform for
exploring new states of matter. In particular, transition metal dichalcogenides
exhibit strong electron-phonon coupling and electron correlations
with a rich phase diagram of charge density wave (CDW) phases. These
materials have been under intense investigations in recent years as
there are many external parameters to control the CDW states, such
as hydro-static pressure \citep{berthier1976evidence,kusmartseva2009pressure,sipos2008frommott},
strain\citep{gao2018atomicscale,qian2021revealing}, chemical doping\citep{qiao2017mottness},
electrostatic doping\citep{yu2015gatetunable}, and intercalation\citep{chatterjee2015emergence,meyer1975properties,morosan2006superconductivity}.
Recently realized control over these states with optical\citep{stojchevska2014ultrafast}
and electrical pulses\citep{mraz2022chargeconfiguration,vaskivskyi2016fastelectronic,yoshida2015memristive}
increased interest in these materials as it opened the way to possible
technological applications.

A particularly interesting transition metal dichalcogenide is the
2$H$-NbSe$_{2}$ in which an incommensurate 3x3 CDW state forms below
$T_{\mathrm{CDW}}=33$ K\citep{moncton1975studyof}. The magnitude
of the CDW energy gap is up to $\sim5$ meV and is wave-vector dependent\citep{rahn2012gapsand}.
Various stripe CDW orders appear with strain\citep{gao2018atomicscale},
intercalation\citep{mogami2021appearance}, or after applying an electrical
pulse\citep{bischoff2017nanoscale}. While in many CDW materials,
the superconducting (SC) state emerges only after applying pressure\citep{kusmartseva2009pressure,sipos2008frommott},
doping\citep{liu2013superconductivity}, or intercalation\citep{morosan2006superconductivity},
2\emph{H}-NbSe$_{2}$ is one of the rare examples in which the CDW
and superconductivity coexist in a pristine sample. Under high pressure,
the CDW is suppressed, while superconducting critical temperature
shows an increase\citep{berthier1976evidence}.

In a typical superconductor, the amplitude (Higgs) mode is not directly
observable as it is weakly coupled to electromagnetic fields and is
over-damped\citep{littlewood1982amplitude,pekker2015amplitudehiggs}.
In 2\emph{H}-NbSe$_{2}$, however, the coupling between superconductivity
and CDW gives rise to a spectroscopically visible SC Higgs mode that
has been observed with Raman spectroscopy\citep{grasset2018higgsmode,measson2014amplitude,sooryakumar1980ramanscattering}.

Optical pump excitation with ultrashort laser pulses can both, excite
collective modes and give additional information on the nature of
electronic states by tracing the single particle and collective mode
dynamics. So far, there have been two investigations of 2\emph{H}-NbSe$_{2}$
with ultrafast pump-probe spectroscopy.\citep{anikin2020ultrafast,payne2020lattice}
\citet{anikin2020ultrafast} investigated 2\emph{H}-NbSe$_{2}$ response
to a high fluence laser excitation in both normal and superconducting
states. \citet{payne2020lattice} investigated a weaker excitation
regime in the CDW state only using a broadband probe. The relatively
short timescale data suggested that the decay time of the laser-excited
CDW state is diverging around the CDW transition temperature. The
critical fluence for the suppression of the CDW state at low-$T$
was estimated to be around 60 $\mu$J/cm$^{2}$.

The observed transient reflectivity signal from recent pump-probe
experiments has, however, proved difficult to interpret as there is
a significant fast electron background signal overlapping with the
CDW response. Additionally, the transient response of a CDW in 2\emph{H}-NbSe$_{2}$
is very long-lived, so a study with longer delay between the pump
and probe would be useful to understand the slow dynamics.

We performed all-optical pump-probe experiments at the 1.55 eV pump-
and probe-photon energy that was not covered in the previous experiments,
and observe a strong charge density wave response with minimal background.
We observe that the CDW state is fragile as it is suppressed at a
laser fluence of the order of $\sim10$ uJ/cm$^{2}$ only, depending
on the pump photon energy. The suppression is not strongly non-thermal
as a large number of phonon degrees of freedom are excited concurrently.
For long pump-probe delays we observe that the excited state decay
time is fluence- and temperature- dependent, where the dynamics is
governed mainly by phonon diffusion processes. We also extend the
previous works by using a larger pump-photon energy (3.1 eV) revealing
a hitherto unreported high energy quasiparticle bottleneck.

At a very low pump fluence, we observe a transient reflectivity response
component due to the superconducting state. Similar to previous pump-probe
experiments we do not observe any coherent oscillations that could
be attributed to the excitation of the SC Higgs mode, or the CDW amplitude
mode.

\begin{figure}
\includegraphics[width=1\columnwidth]{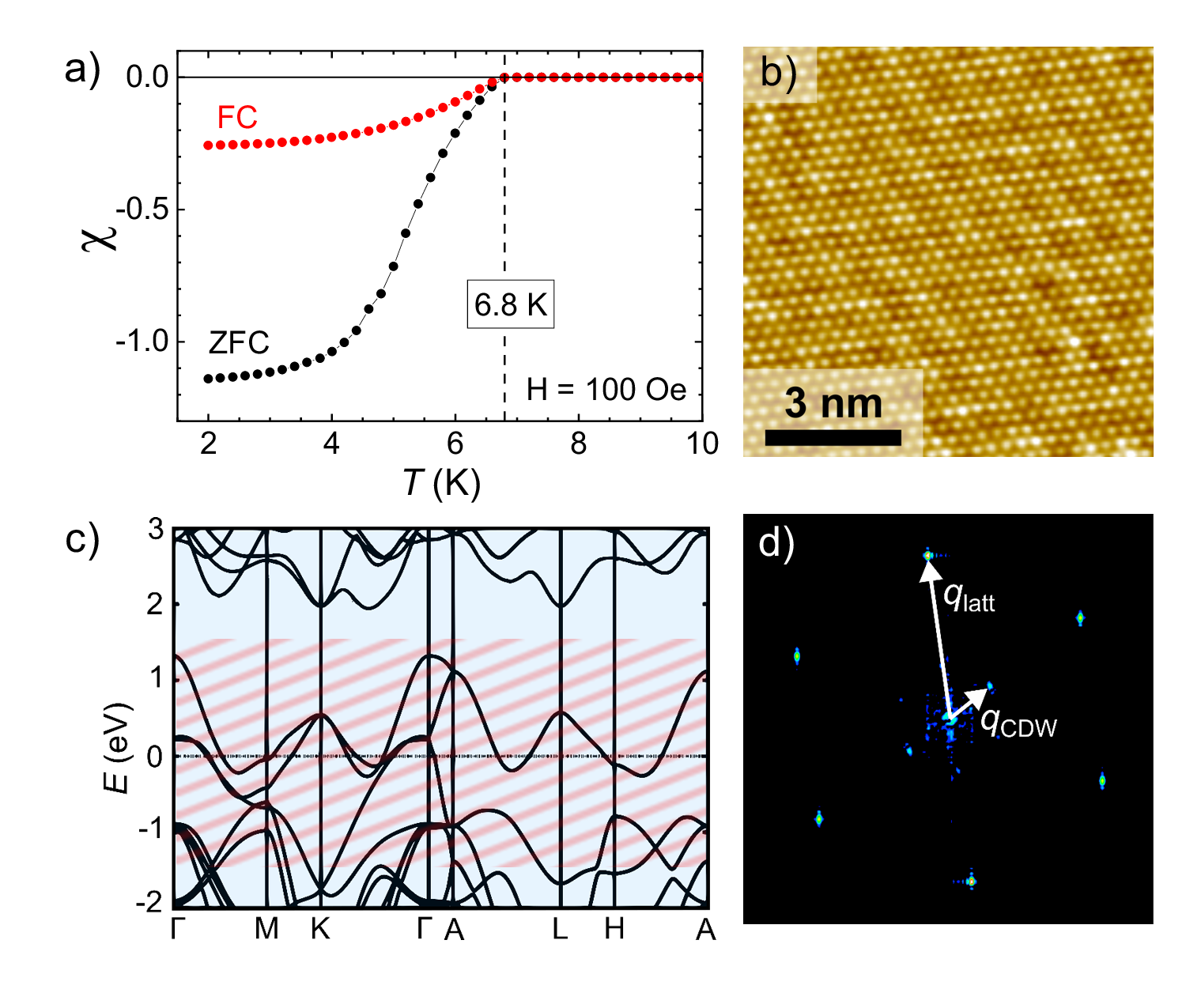}\caption{Sample characterization. a) Temperature dependence of the SQUID magnetic
susceptibility shows the crystals are superconducting below 6.8 K.
b) A 4.2-K scanning tunneling microscope image (set point parameters:
tip bias $V=$ -50 mV, $I=$ 160 pA) and d) the corresponding Fourier
transform showing the $\sim3\times3$ CDW state. c) The electronic
band structure adapted from Ref. {[}\onlinecite{anikin2020ultrafast}{]}.
The red hatched region corresponds to the 1.55 eV photon accessible
excitation range. The 3.1 eV photon accessible excitation range exceeds
the plotted energy range.\label{fig:Sample-characterization}}
\end{figure}

\begin{figure*}
\includegraphics[width=0.9\textwidth]{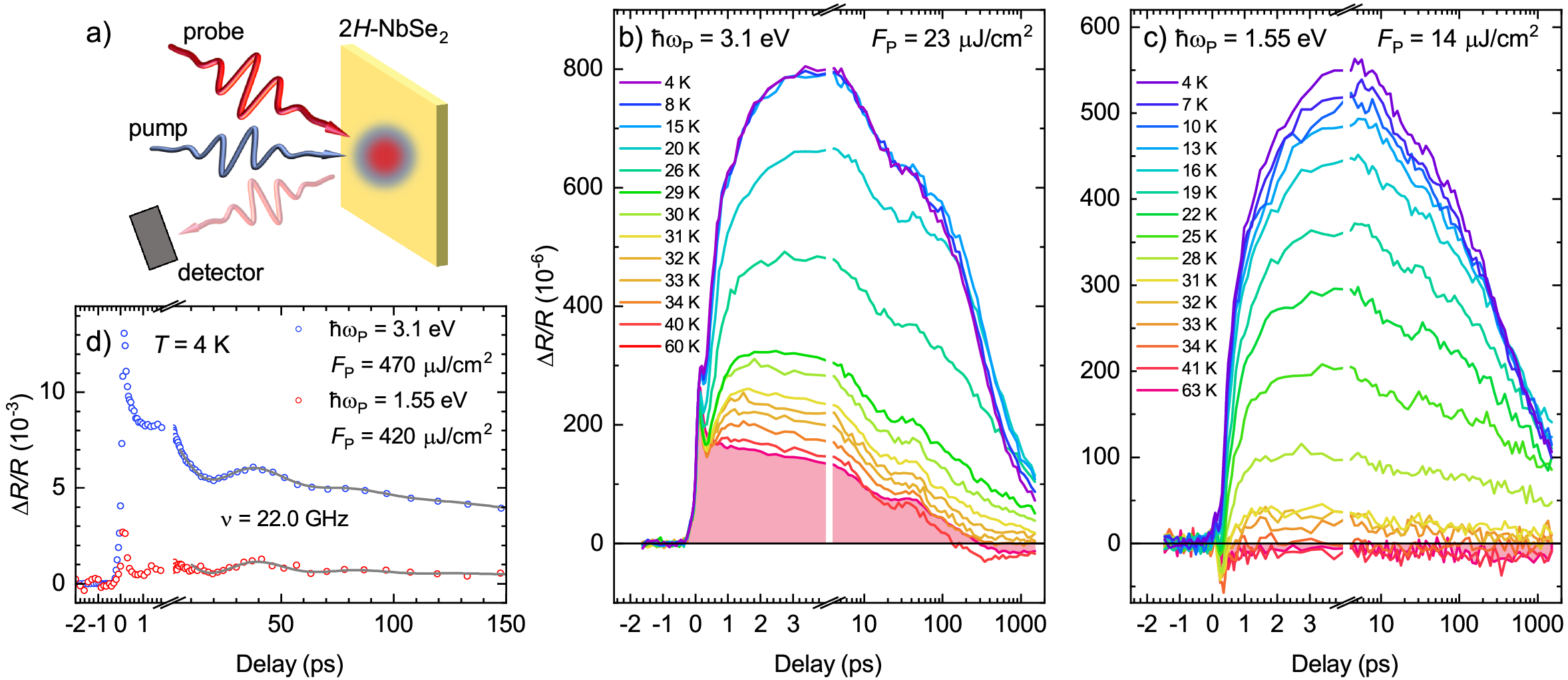}\caption{Temperature dependent transient reflectivity. a) Schematics of the
2-pulse pump-probe experiment. b) and c) $T$-dependent reflectivity
transients with 3.1 eV and 1.55 eV pump-photon energy, respectively.
Note the logarithmic scale after breaks in b) and c). d) High-$F_{\mathrm{P}}$
transients reveal more clearly the coherent sound wave oscillations.
The lines are fits discussed in the text.}
\label{fig:T-dependence}
\end{figure*}

\begin{figure*}
\includegraphics[width=0.9\textwidth]{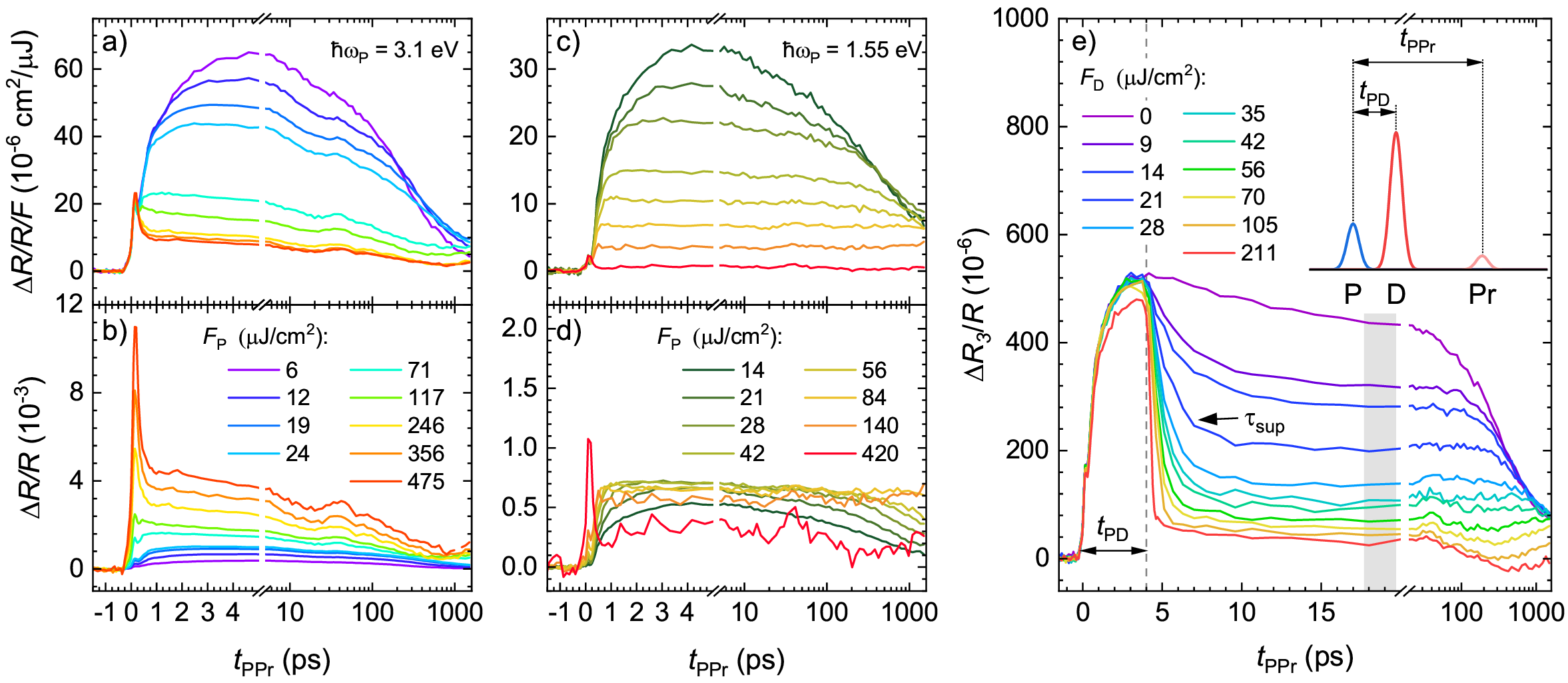}\caption{Fluence dependent transient reflectivity at $T=4$ K. a) $T$-dependent
$F_{\mathrm{P}}$-normalized reflectivity transients at the 3.1 eV
pump-photon energy. Note that curve overlap indicates a linear scaling
with $F_{\mathrm{P}}$. b) The corresponding unnormalized transients.
c), d) The same at 1.55 eV pump-photon energy, respectively. The increased
noise at high $F_{\mathrm{P}}$ is due to the pump scattering. e)
3-pulse transients as a function of the driving-pulse fluence, $F_{\mathrm{D}}$,
at $F_{\mathrm{P}}=19\,\mu$J/cm$^{2}$. The schematics indicating
the pulse sequence is shown in the inset. The shaded region corresponds
to the amplitude readout pump-probe delay discussed in the text while
the vertical dashed line represents the D-pulse arrival time.}
\label{fig:fluence}
\end{figure*}

\section{methods}

\subsection{Sample growth and characterization}

2$H$-NbSe\textsubscript{2} single crystals were synthesized by means
of the chemical vapor transport method from stoichiometric amounts
of niobium foil and selenium powder with iodine as a transport agent.
The material was sealed in a quartz ampule, put into a three-zone
tube furnace with temperature gradient T\textsubscript{H}= 750 C
and T\textsubscript{L} = 680 C, and slowly cooled to room temperature.
Crystal structure and composition were verified with single-crystal
X-ray diffraction and energy-dispersion spectroscopy, respectively.
We used SQUID measurements to determine the magnetic susceptibility
$\chi$. The data show {[}Fig. \ref{fig:Sample-characterization}
a) {]} that the sample is superconducting below 6.8 K. Additionally
we performed an STM characterization of a 2$H$-NbSe\textsubscript{2}
cleaved surface at 4.2 K and observed the $\sim3\times3$ charge density
wave shown in Fig. \ref{fig:Sample-characterization}.

For optical measurements the crystals were exfoliated before mounting
into an optical cryostat to obtain a high-quality surface.

\subsection{Transient optical spectroscopy}

The 2-pulse and 3-pulse transient reflectivity measurements \citep{yusupov2010coherent,naseska2018ultrafast}
were performed using 50-fs linearly polarized regenerative amplifier
pulses at 800 nm wavelength and the $250$ kHz repetition rate. We
used the pump (P) pulses at either the laser fundamental ($\hbar\omega=1.55$~eV)
or doubled ($\hbar\omega=3.1$~eV) photon energy (PE) and the probe
(Pr) pulses at $\hbar\omega=1.55$~eV. In the 3-pulse case we used
additional intense driving (D) pulses (also at $\hbar\omega=1.55$~eV)
with a variable delay with respect to the P pulses (see Fig. \ref{fig:fluence}).

Both the 2-pulse and 3-pulse transient reflectivity, $\Delta R/R$
and $\Delta R_{3}/R$, respectively, were measured by monitoring the
intensity of the weak Pr beam. The large direct contribution of the
unchopped D beam to the total transient reflectivity, $\Delta R$,
was rejected by means of a lock-in synchronized to the chopper that
modulated the intensity of the P beam only. Due to the chopping scheme,
the measured quantity in the 3-pulse experiments is the difference
between the transient reflectivity in the presence of P and D pulses,
$\Delta R_{\mathrm{DP}}(t_{\mathrm{Pr}},t_{\mathrm{P}},t_{\mathrm{D}})$,
and the transient reflectivity in the presence of the D pulse only,
$\Delta R_{\mathrm{D}}(t_{\mathrm{Pr}},t_{\mathrm{D}})$: 
\begin{eqnarray}
\Delta R_{3}(t_{\mathrm{Pr}},t_{\mathrm{P}},t_{\mathrm{D}}) & = & \Delta R_{\mathrm{DP}}(t_{\mathrm{Pr}},t_{\mathrm{P}},t_{\mathrm{D}})\nonumber \\
 &  & -\Delta R_{\mathrm{D}}(t_{\mathrm{Pr}},t_{\mathrm{D}}),\label{eq:DR3}
\end{eqnarray}
where $t_{\mathrm{Pr}}$, $t_{\mathrm{P}}$ and $t_{\mathrm{D}}$
correspond to the Pr, P and D pulse arrival times, respectively. In
the limit of vanishing D pulse fluence $\Delta R_{3}/R$ is reduced
to the standard two-pulse transient reflectivity $\Delta R/R$.

The P/D and Pr beam diameters were 40-70 and 18-30 $\mu$m, respectively.
The Pr fluence was $\sim10$~$\mu$J/cm$^{2}$. For the 3-pulse measurements
the fluence of the P pulse, $\mathcal{F_{\mathrm{P}}}\lesssim20$~$\mu$J/cm$^{2}$,
was kept close to the linear response region. The polarizations of
the P and D beams were perpendicular to the Pr beam polarization with
a random orientation with respect to the crystal axes.

\section{Results}

In Fig. \ref{fig:T-dependence} b) and c) we plot the temperature
dependence of the transient reflectivity at 3.1 eV and 1.55 eV pump-photon
energies (PPE). At the 3.1 eV PPE {[}Fig. \ref{fig:T-dependence}
b){]} the normal-state $\Delta R/R$ consists of the initial sub-picosecond
component followed by a long-lived, slightly oscillatory, response
extending beyond $\sim100$ ps. An additional component with $\sim1$
ps rise-time and a few hundred ps decay time appears on top of the
normal-state response as the temperature is lowered below $T_{\mathrm{CDW}}$.
At the 1.55 eV PPE {[}Fig. \ref{fig:T-dependence} c){]} we observe
a much weaker transient signal above $T_{\mathrm{CDW}}$. Below $T_{\mathrm{CDW}}$,
identically to the 3.1 eV PPE case, a long-lived CDW component emerges
that slowly saturates in amplitude as the temperature is lowered.

In Fig. \ref{fig:T-dependence} d) we show the high-$F_{\mathrm{P}}$
$\Delta R/R$ at both PPE, which more clearly reveals the coherent
damped oscillatory part of the response, which is present also in
the low-$F_{\mathrm{p}}$ transients. By fitting a damped cosine function
(gray curve) we determine the frequency of 22 GHz. The coherent oscillatory
component does not show a notable $T$ dependence at any $F_{\mathrm{P}}$.

Fig. \ref{fig:fluence} a)-d) shows the low-$T$ $F_{\mathrm{P}}$-dependent
$\Delta R/R$ at both PPE. The CDW component shows saturation with
increasing $F_{\mathrm{P}}$ above $\sim20$ $\mu$J/cm$^{2}$. At
3.1 eV PPE the normal-state response, including the oscillatory component,
scales linearly with $F_{\mathrm{P}}$ and overwhelms the transients
at higher $F_{\mathrm{P}}$.

With increasing $F_{\mathrm{P}}$ the CDW component saturates and
develops a flat-top shape with a rather sharp sub-ps rise, while the
relaxation timescale shifts beyond nanosecond timescale. The saturation
behavior is associated with a CDW suppression\citep{stojchevska2011mechanisms}
by the pump pulse. To study the saturation $F_{\mathrm{P}}$ region
avoiding the normal-state background we performed also a 3-pulse experiment\citep{naseska2018ultrafast}
where another strong driving (D) pulse at 1.55 eV PE is introduced
to suppress the CDW independently of the pump pulse. The D pulse is
applied at $t_{\mathrm{DP}}=4$ ps, near the temporal maximum of the
CDW component. As shown in Fig. \ref{fig:fluence} (c) we observe
a suppression\footnote{In the 3-pulse experiment the D pulse is not-chopped an therefore
its contribution to the Pr is suppressed {[}see Eq. (\ref{eq:DR3}){]}.} of the CDW component on a $\tau_{\mathrm{sup}}\sim2.5$ ps timescale
at low $F_{\mathrm{D}}$, which decreases with increasing $F_{\mathrm{D}}$
to, $\tau_{\mathrm{sup}}\sim0.5$ ps, at the complete suppression
{[}see Fig. \ref{fig:fluence} (e){]}.

All high $F$ experiments displayed reversible behavior with the maximum
$F$ being limited to $\sim1$ mJ/cm$^{2}$.

\section{Discussion}

\subsection{Normal state response}

We start by discussing the $T$ and $F$ independent coherent oscillatory
component that corresponds to the photo-excitation induced sound wave\citep{thomsen1986picosecond}.
In order to observe such a wave the real part of the refractive index
must be larger than the imaginary part, $n>\kappa$, at the Pr photon
energy {[}see Appendix \ref{Appendix:sat-mod}, Eq. (\ref{eq:mov-mir}){]}.
Calculating the optical constants at the Pr photon energy (1.55 eV)
from \citealt{dordevic2001anisotropic} we obtain the out-of-plane
sound group velocity, $c_{\mathrm{s}z}=2540\pm70$ m/s.\footnote{The error bar is the non-linear least square fit error and does not
take into account the index of refraction error.} On the other hand, taking the literature static elastic constant\citep{feldman1976elastic},
$c_{33}=46$ GPa, we obtain $c_{\mathrm{s}z}=2670$ m/s. The $\sim10\%$
discrepancy can be attributed to the experimental inaccuracy of the
refraction index.

Next, we discuss the PPE dependence of the normal state $\Delta R/R$
that is much larger at 3.1 PPE. The band structure of $2H$-NbSe$_{2}$
is characterized by a band gap between the partially occupied bands,
with mixed Se-$p$/Nb\textendash $d$ character extending up to $\sim1.2$
eV above the Fermi energy, and, separated by a gap, the Nb\textendash $d$
character dominated bands starting $\sim$2 eV above the Fermi energy\citep{krasovskii2002unoccupied}
{[}see Fig. \ref{fig:Sample-characterization} c){]}. The photo excitation
of the latter is possible only at the 3.1 eV PPE and can contribute
to the 1.55 eV $\Delta R/R$ only while they are occupied. The 3.1
eV PPE transient response extending beyond 100 ps at low $F$ therefore
suggests that a slowly-relaxing weakly-coupled electron ``pocket''
exists in the bands $\sim2$ eV above the Fermi energy. This is somewhat
unexpected since $2H$-NbSe$_{2}$ is metallic at any $T$ and the
Auger relaxation across the band gap is not \emph{a priori} forbidden
by energy conservation.

\subsection{CDW and SC state response}

We turn first to the CDW component. With decreasing $T$ it appears
below $T_{\mathrm{CDW}}$ {[}Fig. \ref{fig:T-dependence} b), c){]},
similar as reported previously using the 1.03 eV Pr photon energy,\citep{payne2020lattice}
but shows different low-$T$ amplitude saturation. In Ref. {[}\onlinecite{payne2020lattice}{]}
it saturates below $\sim25$ K, while in the present case, using the
1.55 eV Pr photon energy, it clearly saturates only below $T\sim15$
K at 3.1 eV PPE {[}Fig. \ref{fig:T-dependence} b){]}, while at 1.55
eV PPE it does not saturate down to the lowest $T=4$ K {[}Fig. \ref{fig:T-dependence}
c){]}. The saturation seems to be connected with the excitation density,
which was the lowest, $F_{\mathrm{P}}=14\:\mu$J/cm$^{2}$, in the
present 1.55 eV PPE experiment. We therefore need to discuss the excitation
density dependence before continuing to discuss the $T$ dependence.

\subsubsection{Excitation density dependence in the CDW state}

\begin{figure}
\includegraphics[width=0.5\columnwidth]{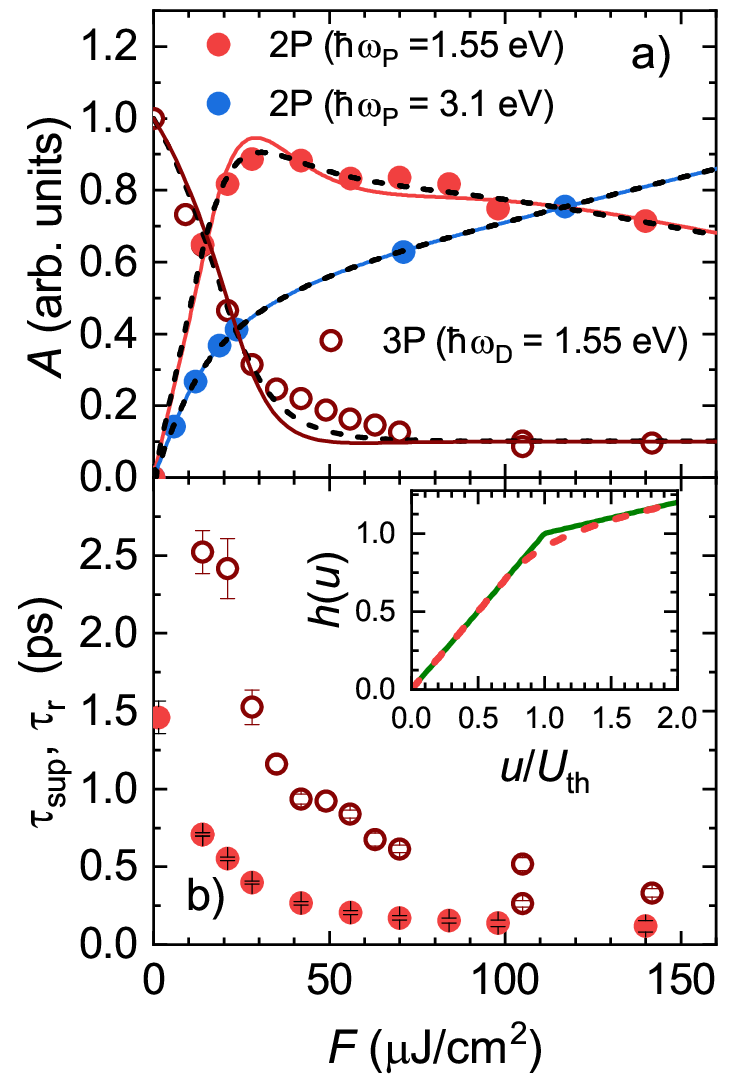}

\caption{Selected $F$-dependent transient response parameters at $T=4$ K.
a) The fluence dependence of the signal amplitudes. The open symbols
correspond to the 3-pulse experiment. The lines are the saturation
model fits discussed in text. b) The rise time (full circles) and
3-pulse signal suppression time (open circles) as function of $F$.
The inset to b) shows the assumed saturation functions discussed in
text.\label{fig:Selected-F-dependent}}
\end{figure}

\begin{table*}
\begin{tabular}{cccccccc}
$\hbar\omega$ (eV) & $n$$\mathrm{^{\ref{fn:dordevic}}}$ & $\alpha^{-1}$(nm)\footnote{Calculated from data in Ref. [\onlinecite{dordevic2001anisotropic}].\label{fn:dordevic}} & $R$$\mathrm{^{\ref{fn:dordevic}}}$ & $\phi$ & $F_{\mathrm{th}}$ ($\mu$J/cm$^{2}$) & $U_{\mathrm{th}}$ (J/cm$^{3}$) & $T_{\mathrm{th}}$(K)\tabularnewline
\hline 
\hline 
3.1 (2-pulse) & 3.5 & 21 & 0.38 & \multirow{3}{*}{1.57\footnote{The phase shift {[}Eq. \eqref{eq:phi}{]} value was fit for the case
of the 1.55 eV PPE 2-pulse experiment and taken as a fixed parameter
for the 3.1 eV PPE 2-pulse and 3-pulse experiments.}} & $4.5\pm0.7$ & $1.3\pm0.2$ & $\sim28$\tabularnewline
1.55 (2-pulse) & 3.2 & 83 & 0.3 &  & $14\pm0.5$ & $1.2\pm0.05$ & $\sim26$\tabularnewline
1.55 (3-pulse) & 3.2 & 83 & 0.3 &  & $19\pm0.8$ & $1.6\pm0.07$ & $\sim29$\tabularnewline
\end{tabular}\caption{The CDW destruction threshold fluences, energy densities, estimated
peak transient temperatures at $F_{\mathrm{th}}$, and the relevant
optical parameters.\label{tab:The-CDW-destruction}}
\end{table*}

In Fig. \ref{fig:Selected-F-dependent} a) we summarize the $F$ dependence
experiments from Fig. \ref{fig:fluence} by plotting the magnitudes\footnote{In the case of the 3-pulse experiment we define the magnitude as value
of the transient reflectivity at $t_{\mathrm{PPr}}\sim18$ ps {[}the
gray bar in Fig. \ref{fig:fluence} c){]} where a plateau is observed
in the presence of the D pulse..} of the transient responses as functions of the relevant fluence for
the standard two-pulse 1.55 eV and 3.1 eV PPE experiments as well
as the 3-pulse experiment. All the magnitudes show (partial) saturation
with increasing $F$ that is attributed to the CDW state suppression\citep{stojchevska2011mechanisms},
with an onset around $F\sim10$ $\mu$J/cm$^{2}$. The virtually linear
$F$ dependence beyond the saturation region is the consequence of
the normal-state background, which scales linearly with $F$ and is
the largest in the case of the 3.1 eV PPE.

To extract the absorbed-energy density at the CDW-suppression threshold,
$U_{\mathrm{th}}$, we apply the saturation model\citep{kusar2008controlled,naseska2018ultrafast}.
In the model we assume a nonlinear transient dielectric function,
$\Delta\epsilon$, dependence on the absorbed-energy density, $U$,
at $U_{\mathrm{th}}$ {[}Appendix \ref{Appendix:sat-mod}, Eq. \eqref{eq:sat}{]}.
In addition, we take into account the inherent excitation density
inhomogeneity and the optical-probe response kernel\citep{naseska2018ultrafast}
{[}Appendix \ref{Appendix:sat-mod}, Eq. \eqref{eq:sat-dr}{]}. The
fits to the $F$-dependent $\Delta R/R$ magnitudes, derived from
the model, as described in detail in Appendix \ref{Appendix:sat-mod},
are shown in Fig. \ref{fig:Selected-F-dependent} a).

\begin{figure}
\includegraphics[width=0.8\columnwidth]{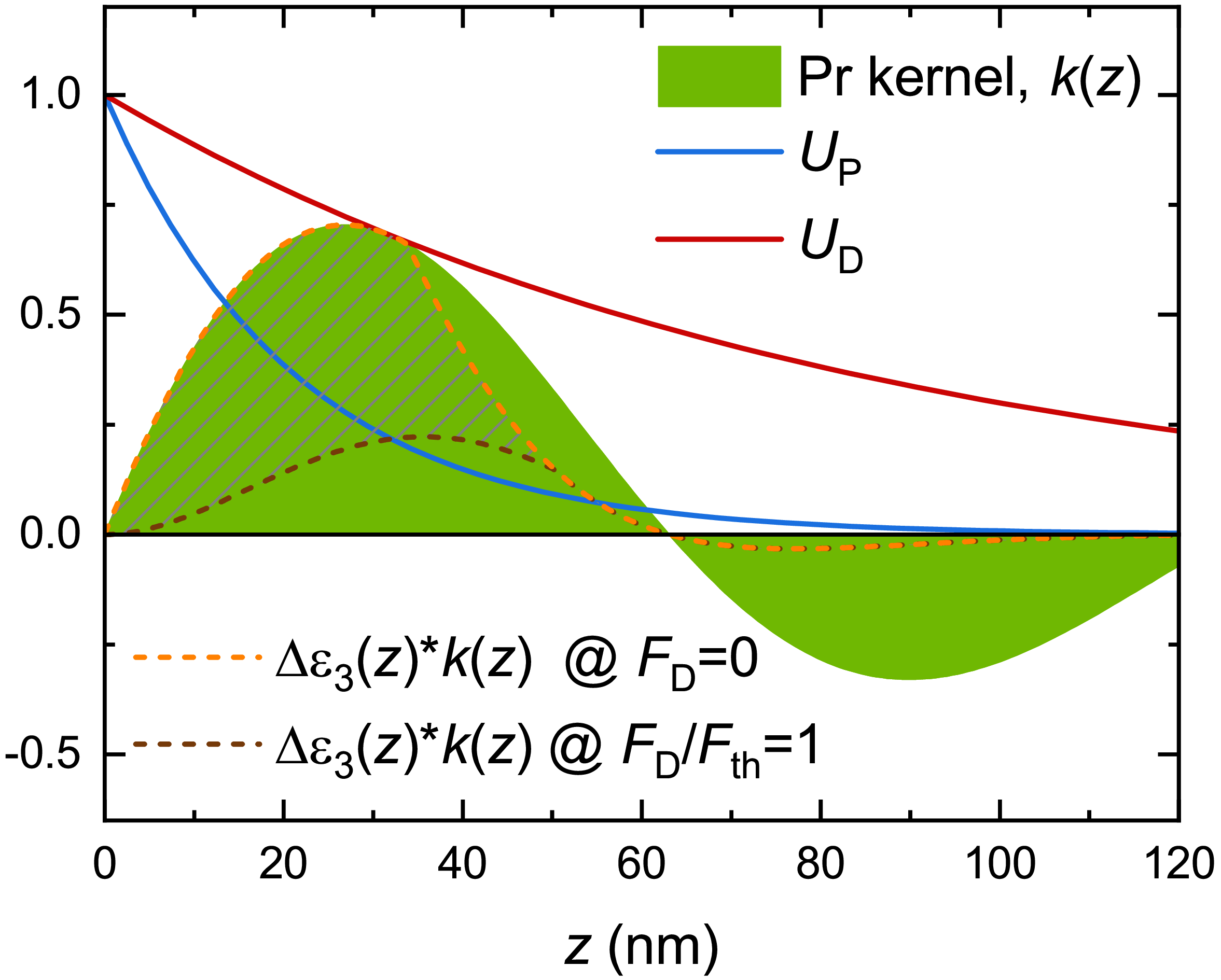}\caption{Depth dependence of the probe kernel {[}Eq. \eqref{eq:sat-dr}{]}
in comparison to the normalized P and D deposited energy density profiles.
The model kernel-weighted 3-pulse response {[}Eq. \eqref{eq:sat3p}{]}
at two selected D fluences is also shown with the dashed lines. The
hatched region corresponds to the 3-pulse signal suppression profile
at $F_{\mathrm{D}}=F_{\mathrm{th}}$. The kernel parameters were taken
from Tab. \ref{tab:The-CDW-destruction}. The phase shift, $\phi\sim\pi/2$,
obtained from the fit is such, that the negative-sign kernel region
is relatively deep in the sample, minimizing its contribution to the
signal.\label{fig:kernel}}
\end{figure}

The saturation model\citep{naseska2018ultrafast} predicts a possibility
of oscillatory transient-response-amplitude $F$ dependence, when
the probe-PE refraction-index real part, $n$, exceeds the refraction-index
imaginary part, $\kappa$, (see Appendix \ref{Appendix:sat-mod} and
Ref. {[}\onlinecite{naseska2018ultrafast}{]} ) due to the interference
effect on a thin surface layer, where the CDW order parameter (OP)
is optically suppressed. Despite $n/\kappa\approx4.1$ in $2H$-NbSe$_{2}$,\footnote{At the 1.55 eV Pr photon energy\citep{dordevic2001anisotropic}.}
we observe no clear oscillatory behavior in the data. This can be
partially attributed to the lateral thickness distribution of the
suppressed-OP layer due to the Gaussian beam profile and a particular
combination of the static and transient dielectric function components
{[}see Eq. (\ref{eq:phi}){]} resulting in the kernel shape (see Fig.
\ref{fig:kernel}) that minimizes the oscillations.

In the case of the 1.55 eV PPE 2-pulse experiment, however, the model
oscillations can not be completely washed out by the above mentioned
effects effects, as indicated by the red curve in Fig. \ref{fig:Selected-F-dependent}
a). Assuming some smearing in the nonlinear local dielectric function
$U$-dependence around $U_{\mathrm{th}}$ {[}see Eq. \eqref{eq:sat}
and insert to Fig. \ref{fig:Selected-F-dependent} b){]} leads to
virtually complete suppression of the oscillations as indicated by
the black dashed lines in Fig. \ref{fig:Selected-F-dependent} d).
The smearing suggests an absence of a sharp boundary between the suppressed
and non-suppressed OP regions which is plausible due to presence of
a strong transient chemical potential gradient, causing quasiparticle
and possibly CDW-sliding currents between the regions.

The application of smearing does not significantly affect the obtained
$U_{\mathrm{th}}$ so the values from the fits in the absence of the
smearing are reported in Tab. \ref{tab:The-CDW-destruction}. Focusing
first on the 2-pulse experiments, $U_{\mathrm{th}}$ is found somewhat
larger at the 3.1 eV PPE, but the difference is within the fitting
error bars. Using the heat capacity data\citep{harper1975heatcapacity}
we estimate the enthalpy change when heating the sample thermally
to $T_{\mathrm{CDW}}$ to be, $\Delta H{}_{\mathrm{CDW}}\thickapprox2.5$
J/cm$^{3}$. The experimental $U_{\mathrm{th}}$ values around a half
of $\Delta H{}_{\mathrm{CDW}}$ therefore indicate that the CDW suppression
is non thermal, however, due to the small low-$T$ phonon specific
heat capacity the peak transient temperature, $T_{\mathrm{th}}$,
corresponding to the fully thermalized excitation volume is relatively
large, above $\sim26$\,K at $U_{\mathrm{th}}$ (see Tab. \ref{tab:The-CDW-destruction}).

Applying the saturation model to the 3-pulse experiment (Eq. \ref{eq:sat3p})
results in a larger $U_{\mathrm{th}}$ and the fit departs from the
data at higher $F_{\mathrm{D}}$ {[}see Fig. \ref{fig:Selected-F-dependent}
a){]} even when assuming the smearing (see the dashed fit curve).
To understand this one has to take into account important differences
between the 2-pulse and 3-pulse experiments. (i) At the used experimental
parameters, the 3.1 eV PE P pulse, with $F_{\mathrm{P}}/F_{\mathrm{th}}\sim5$,
suppresses the CDW within the depth, $z_{\mathrm{P}}\sim34$\,nm,
before the arrival of the D pulse. (ii) The measured $\Delta R_{3}$
is referenced to $\Delta R_{\mathrm{D}}$, obtained in the presence
of the D-pulse only {[}Eq. \eqref{eq:DR3}{]}. (iii) The $\Delta R_{3}$
amplitude is read out $\sim14$~ps ($\sim18$~ps) after the D (P)
pulse arrival {[}the gray band in Fig. \ref{fig:fluence} e){]}.

Due to (i), (ii) and Eq. \eqref{eq:sat3p} the 3-pulse signal at any
$F_{\mathrm{D}}$ depends on the detailed behavior of $\Delta\epsilon$
near the saturation threshold as most of the signal comes from the
P-suppressed region (see also Fig. \ref{fig:kernel}). Moreover, due
to (iii) most likely some thermalization already takes place at the
readout $t_{\mathrm{PPr}}$. The measured 3-pulse saturation therefore
corresponds to the properties of a partially thermalized strongly
excited material even at low $F_{\mathrm{D}}$, which cannot be taken
into account by the present simple model.

Comparing to superconductors and, in particular, to the large-gap
CDWs\citep{stojchevska2011mechanisms}, a significantly larger amount
of the absorbed energy is lost to the phonon bath. In the large-gap
CDWs, where the gap energy, $2\Delta$, exceeds the maximum phonon
energy, the phonons do not appear to be strongly involved in the OP
suppression at all.\citep{stojchevska2011mechanisms} $U_{\mathrm{th}}$
is comparable to the CDW condensation energy and the suppression is
highly non-thermal irrespective to the presence of ungapped Fermi
surface. This suggests that the initial high-energy photo-excited
quasiparticle relaxation is dominated by Auger processes that excite
the quasiparticles across the CDW gap, and the CDW is suppressed before
the energy is transferred to phonons.

In superconductors, where $2\Delta$ falls within the phonon energy
spectrum, it was argued\citep{stojchevska2011mechanisms} that the
initial high-energy photo-excited quasiparticle relaxation must be
dominated by phonons since the data indicate that most of the absorbed
optical energy is lost to the sub-gap phonons that cannot contribute
to the pair breaking, and consequently to the superconducting OP suppression.
Despite this, $U_{\mathrm{th}}$ is found much smaller than the enthalpy
change when heating the sample thermally to $T_{\mathrm{c}}$, so
the phonon population must remain highly non-thermal on the picosecond
suppression timescale.

In $2H$-NbSe$_{2}$, the CDW gap is in the $\sim1-6$ meV range on
the K pockets with negligible/zero CDW gap on the $\Gamma$ pockets\citep{rahn2012gapsand}
while the phonon spectrum extends to\citep{murphy2005phononmodes}
$\sim30$ meV. The energy scales are therefore somewhat similar to
the lower-$T_{\mathrm{c}}$ cuprate superconductors\citep{stojchevska2011mechanisms}.
The relative amount of the gapless electronic density of states is,
however, larger in the present case \footnote{The CDW gap is concentrated on few hot spots with large amounts of
ungapped Fermi surface.\citep{borisenko2009twoenergy,rahn2012gapsand}} and might be instrumental in transferring the absorbed optical energy
to the sub-gap-energy degrees of freedom that are inefficient for
the CDW suppression. The ungapped inner-most $\Gamma$ pocket, which
shows strong coupling to the highest energy phonons,\citep{rahn2012gapsand}
stands out in particular, as a possible relaxation channel that bypasses
the energy relaxation through the CDW-gapped Fermi surface regions.

The CDW OP suppression times behave quite differently in the 2-pulse
and 3-pulse experiments. The 2-pulse transient reflectivity onset
in CDWs is often dominated by the coherent amplitude mode (AM) and
phonon excitation\citep{demsar1999singleparticle,yusupov2008singleparticle,stojchevska2014ultrafast,stojchevska2017evolution,nasretdinova2019timeresolved,kuo2019characterization}
with the rise time of $\sim\nicefrac{1}{4}$ of the AM period. The
low-$T$ AM mode frequency in 2\emph{H}-NbSe$_{2}$ is\citep{mialitsin2011fanoline}
$\sim40$ cm$^{-1}$ so the corresponding, $\tau_{\mathrm{r}}\sim0.2$
ps, rise time would be much shorter than the present low-$F$ data
with, $\tau_{\mathrm{r}}\sim1.5$ ps, show.

The behavior is therefore more similar to the case of the conventional
and low-$T_{\mathrm{c}}$ cuprate superconductors\citep{kusar2008controlled,demsar2003pairbreaking,beck2011energygap,akiba2023photoinduced}
where the low-excitation-density quasiparticle population evolves
according to the Rothwarf-Taylor\citep{rothwarf1967measurement,kabanov2005kinetics}
bottleneck model, with the non-equilibrium-phonons dominated\citep{demsar2003pairbreaking,beck2011energygap}
initial conditions. In such case the rise time is in a picoseconds
range and it drops with the excitation density, as in the present
case.

The CDW state in 2\emph{H}-NbSe$_{2}$ is, however, not fully gaped
and the bottleneck\citep{kabanov2005kinetics} is rather weak, as
indicated by the relatively large $U_{\mathrm{th}}$. While the slow
rise-time timescale suggests the non-equilibrium-phonons dominated
suppression it is not clear whether the plain Rothwarf-Taylor model
is applicable to explain the weak-excitation\footnote{At strong excitations the saturation non linearity naturally leads
to the drop of the rise time.} rise-time behavior in the present case. Nevertheless, the CDW gap
on the K pockets seem to somehow avoid fast suppression during the
initial photo-excited carrier relaxation suggesting that the dominant
high-energy relaxation involves mostly phonons and the L-pockets quasiparticles.

Compared to the 2-pulse experiment, the 3-pulse-experiment CDW OP
suppression time {[}Fig. \ref{fig:Selected-F-dependent} b){]} is
significantly longer. As discussed above, the 3-pulse signal at any
$F_{\mathrm{D}}$ reflects the properties of the highly-excited and
strongly-suppressed CDW-OP region, where a critical slowing down of
the dynamics might take place. However, the 2-pulse data do not show
any critical slowing-down of the rise time near $T_{\mathrm{CDW}}$
ruling out the hypothesis, unless the highly excited state phase transition
fundamentally differs from the thermal one.

The slower 3-pulse suppression dynamics therefore remains puzzling.
A possible origin of such dynamics could be formation of a spatially
inhomogeneous OP fluctuating state near $F_{\mathrm{th}}$, consistent
also with the absence of the $\Delta R/R$ amplitude oscillations
in the 2-pulse $F$-dependence.

\subsubsection{Signatures of the SC state}

\begin{figure}
\includegraphics[width=1\columnwidth]{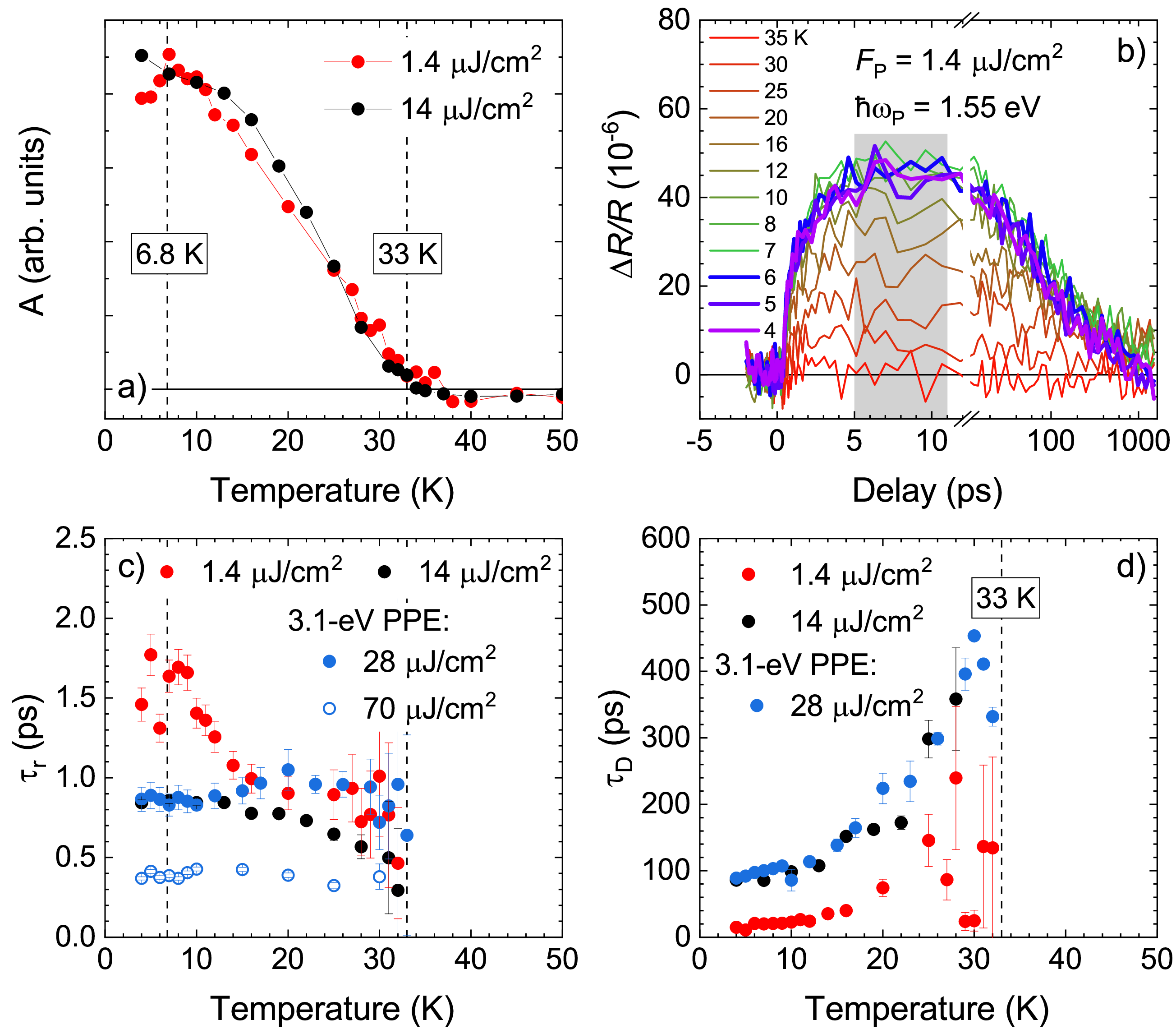}\caption{Selected $T$-dependent transient response parameters. a) The normalized
transient reflectivity amplitude as a function of $T$ at two excitation
fluences at 1.55 eV pump photon energy. b) $T$ dependence of the
lowest pump-fluence transient reflectivity. The shaded region corresponds
to the averaging region of the amplitude readout. c) and d) the rise
time and relaxation time as a function of $T$, respectively.\label{fig:Selected--dependent-parameters.}}
\end{figure}

For both 1.55 eV and 3.1 eV PPE, the transient reflectivity with $F_{\mathrm{P}}$
of tens of $\mu$J/cm$^{2}$, does not show any appreciable change
of signal in the SC state below $T\sim7$ K. However, reducing the
fluence to an extremely low value, $F_{\mathrm{P}}=1.4\:\mu$J/cm$^{2}$,
a slight drop in the amplitude is observed when entering the superconducting
state below $T_{\mathrm{c}}=6.8$ K {[}Fig. \ref{fig:Selected--dependent-parameters.}
a), b){]}. The magnitude of the drop is rather small, and is comparable
to the signal noise level {[}Fig. \ref{fig:Selected--dependent-parameters.}
b){]}. There is also a tiny difference between the CDW and SC-CDW
transients at the longest delays suggesting the presence of a tiny
negative long-lived SC component. The absence of any SC state signature
at $F_{\mathrm{P}}=14$ $\mu$J/cm$^{2}$ response indicates that
the SC induced contribution is already saturated at $F_{\mathrm{P}}=1.4\:\mu$J/cm$^{2}$.
Indeed, the low-$T$ heat capacity is so small that the excited volume
transient $T$ exceeds $T_{\mathrm{c}}$ already around $F_{\mathrm{P}}=0.5\:\mu$J/cm$^{2}$
at 1.55 eV PPE. The possible observation of the optically coherently
excited SC Higgs mode is therefore hindered by the noise level of
the present experiment.

\begin{figure*}
\includegraphics[width=1\textwidth]{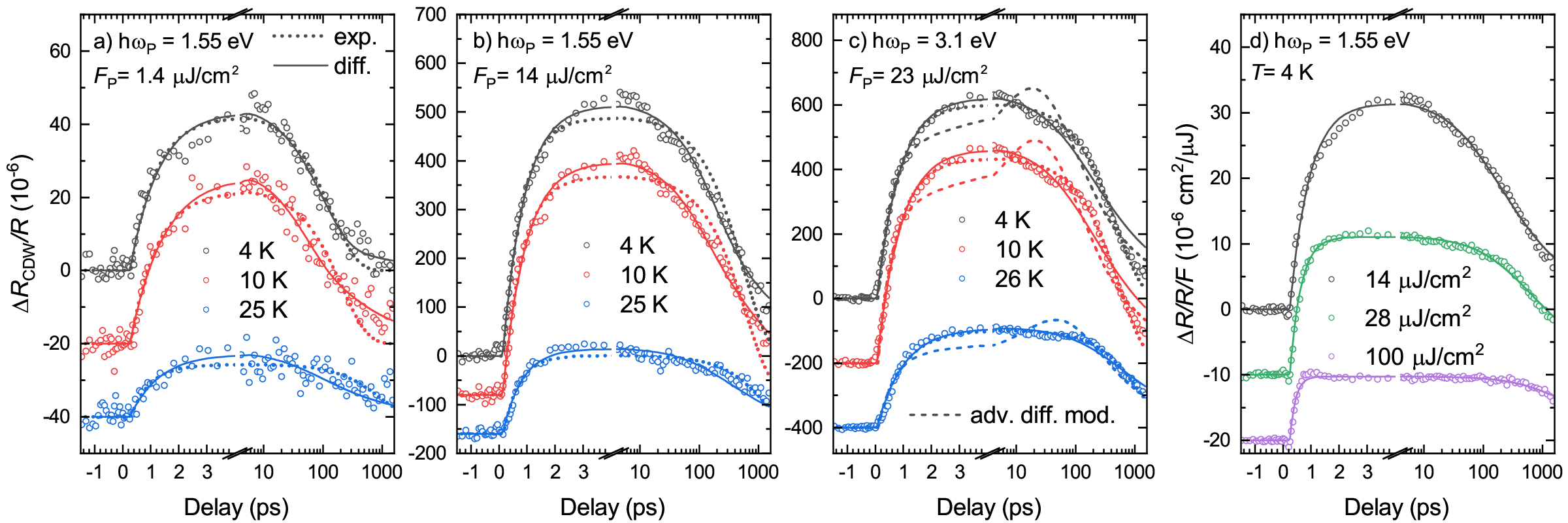}\caption{Examples of fits to the data at different excitation conditions. The
lines correspond to various decay models discussed in text.\label{fig:Examples-of-fits}}
\end{figure*}

\subsubsection{CDW state temporal dynamics}

For discussing the temporal dynamics of the CDW transient reflectivity
component we mostly focus on the 1.55 eV PPE data, where the normal-state
response is negligible and the rise and decay times can be more reliably
quantitatively extracted by simple few parameter fits. A similar analysis
is presented also for the 3.1 eV PPE data. However, in addition to
the larger normal-state response obscuring the CDW dynamics, most
of the 3.1 eV PPE data correspond to the nonlinear response region
because the laser fluences used were larger and the optical penetration
depth, $\alpha^{-1}$, is $\sim4$ times shorter at the 3.1 eV PPE
(see Tab. \ref{tab:The-CDW-destruction}) than at the 1.55 eV PPE.

To remove the oscillatory acoustic component contribution (and the
large normal-state response in the 3.1 eV PPE case) we subtract the
normal state $\Delta R/R$ just above $T_{\mathrm{CDW}}$ before fitting
to obtain the CDW (+ SC) component shown in Fig. \ref{fig:Examples-of-fits}
together with different fits discussed below.

We note that the start of the rise of the CDW component in Fig. \ref{fig:Examples-of-fits}
appears delayed for $\sim200$ fs with respect to the pump pulse as
can be the most clearly seen in Fig. \ref{fig:fluence} d). This delay
can be attributed to the initial relaxation of the high-energy quasiparticles
by Auger and/or phonon emission processes.

While the CDW component rise time dynamics can be described reasonably,
but not to the finest details, using a single exponential function,
the decay dynamics is not exponential (see dotted lines in Fig. \ref{fig:Examples-of-fits}).
It turns out that the 1.55 eV PPE data can be described by a simple
three parameter function assuming a diffusive decay:

\begin{equation}
\Delta R/R=A\left[1/\sqrt{1+t/\tau_{\mathrm{D}}}-\exp\left(-t/\tau_{\mathrm{r}}\right)\right],\label{eq:fit-func}
\end{equation}
in the full CDW temperature range. The first term in the brackets
corresponds to a simplified 1D diffusion dynamics\citep{mertelj2009finestructure},
where the characteristic diffusion time, $\tau_{\mathrm{D}}=\nicefrac{z_{0}^{2}}{4D}$,
corresponds to the optical penetration depth, $z_{0}\sim\alpha^{-1},$
diffusion length-scale. The second term describes the rise time dynamics.

The slight discrepancy of the diffusive-decay fit at long timescales
at the lowest $T$ and 1.55 PPE can be attributed to the $T$-dependence
of the diffusion constant and the presence of the additional SC component.
The much more prominent diffusive-decay fit failure {[}see Fig. \ref{fig:Examples-of-fits}
c){]} at 3.1 eV PPE, where the exponential decay\footnote{Replacing the first term in brackets with $\exp(-t/\tau)$.}
fits are somewhat better, however, cannot be of the same origin.

The $\Delta R/R$ decay dynamics at the 3.1 PPE could be affected
by the factor of $\sim4$ smaller\footnote{With respect to the probe.}
pump optical penetration depth, resulting in the larger inhomogeneity
within the probed volume, and the strong nonlinearity due to the higher
excitation density. In Fig. \ref{fig:Examples-of-fits} c) we therefore
plot also fits (dashed lines) obtained in the framework of the saturation
model (see Appendix \ref{Appendix:sat-mod}) taking into account a
simplified 1D diffusion depth profile \eqref{eq:diff} and the nonlinearity
\eqref{eq:sat} at $U_{\mathrm{th}}$. Unfortunately, the advanced
diffusion model results in a very poor fit at the tens-of-picoseconds
timescale.\footnote{The advanced diffusion model fits virtually overlap the simple diffusion
model at 1.55 eV PPE and are not shown.} In the model the CDW OP suppression is expected to spread deeper
into the sample, initially, leading \footnote{In the framework of the saturation model.}
to the transient reflectivity increase on tens-of-picoseconds timescale
due to the short 3.1 eV PE pump optical penetration depth and the
relatively deeper sensitivity of the probe kernel (Fig. \ref{fig:kernel}).
The absence of such a peak in the data therefore suggests an initial
fast (a few-picosecond) spread of the excitation beyond the optical
penetration depth or/and thermalization of the excited degrees of
freedom to the phonon heat bath on a $\sim10$~ps timescale, which
are not included in the saturation-model diffusion fits.

Moreover, due to a complete CDW OP suppression, formation of topological
defects cannot be excluded.\citep{mertelj2013incoherent,orenstein2023subdiffusive}
This effect should slow-down the recovery and can influence the shape
of the transients on several tens\citep{mertelj2013incoherent} of
picoseconds. Due to the similarity of the timescales it is unfortunately
not possible to disentangle these effects in the current data.

The characteristic diffusion time, $\tau_{\mathrm{D}}$, obtained
from the fits is strongly $F_{\mathrm{P}}$ dependent {[}Fig. \ref{fig:Selected--dependent-parameters.}
d){]}, increasing with increasing fluence from $\sim10$ ps at the
lowest $F_{\mathrm{P}}=1.4$ $\mu$J/cm$^{2}$ to $\sim90$ ps at
$F_{\mathrm{P}}=14$ $\mu$J/cm$^{2}$ at the lowest $T$. The increase
of $\tau_{\mathrm{D}}$ with increasing $F$ can be attributed to
the increased transient $T$ resulting in a decrease of the diffusion
constant.

The low-$T$ weak-excitation out-of-plane diffusion constant, $D_{\mathrm{op}}\sim1$
cm$^{2}$/s, obtained from $\tau_{\mathrm{D}}$, appears $\sim30$
times smaller than the equilibrium in-plane thermal diffusion constant,
$D_{\mathrm{eq-ip}}$.\footnote{Calculating the $T$-dependent in-plane equilibrium heat diffusion
constant from literature data\citep{harper1975heatcapacity,beletskii1998thermal}
we obtain $D_{\mathrm{eq-ip}}\sim1$ cm$^{2}$/s at $T_{\mathrm{CDW}}$
increasing with decreasing $T$ to $D_{eq-\mathrm{ip}}\sim30$ cm$^{2}$/s
at $T_{\mathrm{c}}$.}$^{,}$ \footnote{In the strong excitation cases, $D_{\mathrm{op}}$ is found smaller,
$\sim0.2$ and $\sim0.01$~cm$^{2}$/s (at $T=4$~K) for the 1.55
eV PPE and 3.1 eV PPE excitation, respectively. Here, however, one
should take into account that the transient $T$ in the strong excitation
cases is of the order of $T_{\mathrm{CDW}}$. $D_{\mathrm{op}}$ is
therefore found $\sim20$ times smaller (at $T\sim26$ K) and $\sim40$
times smaller (at $T\sim48$ K) than $D_{\mathrm{eq-ip}}$ for the
1.55 eV PPE and 3.1 eV PPE excitation, respectively.} Taking into account that $\sim\nicefrac{1}{3}$ of the in-plane thermal
conductivity in 2$H$-NbSe$_{2}$ is due to phonons at low $T$,\citep{roeske1977superconducting}
this anisotropy is compatible with the rather large resistivity anisotropy\citep{leblanc2010resistivity,pfalzgraf1987theanisotropy}
of at least several hundred, if the phonon thermal conductivity is
less anisotropic. With $\sim\nicefrac{1}{3}$ of in-plane heat transfer
due to phonons the equilibrium thermal diffusion constant anisotropy
of $\nicefrac{D_{\mathrm{eq-ip}}}{D_{\mathrm{eq-op}}}\sim10$ would
be consistent with the data.

Since the relaxation appears diffusion dominated it is unclear whether
the increase of $\tau_{\mathrm{D}}$ with increasing $T$ reflects
also a slowing down of the non-equilibrium OP relaxation/thermalization.
In the framework of the phenomenological CDW dynamics model by \citet{schaefer2014collective}
the absence of the electronic-mode critical slowing down at $T_{\mathrm{CDW}}$
would suggest the adiabatic dynamics, where the electronic OP instantly
follows the lattice motion. In this limit a strong softening of the
oscillatory AM is expected.\citep{schaefer2014collective} Such mode
has not been observed in the 1.55 eV probe PE transient reflectivity,
presumably due to a low Raman cross-section and/or inefficient coherent
excitation\footnote{While the observed transient reflectivity dynamics clearly contains
some sub-picosecond-timescale components the CDW component rise time
is significantly slower than the AM oscillation period.}. \footnote{\citet{anikin2020ultrafast} observed a $\sim4$-THz ($\sim130$ cm$^{-1}$)
coherent mode at 2.2 eV PPE, which softens by $\sim10\%$ with increasing
$T$ towards $T_{\mathrm{CDW}}$. The data were, however, taken at
rather high $F_{\mathrm{P}}=250\:\mu$J/cm$^{2}$, well above the
CDW destruction threshold fluence, $F_{\mathrm{th}}$, and the mode
persists in the normal state so it cannot correspond to the back-folded
lattice mode contributing to the CDW OP.} The Raman data\citep{mialitsin2011fanoline}, however, indicate a
moderate softening of the low-$T$ 40-cm$^{-1}$ A$_{\mathrm{1g}}$
Raman mode by $\sim10$ cm$^{-1}$ and rather large broadening from
$\gamma\sim20$ cm$^{-1}$ at low $T$ to $\gamma\sim60$ cm$^{-1}$
near $T_{\mathrm{CDW}}$, which suggests the non-adiabatic electronic
order parameter dynamics,\citep{schaefer2014collective} where a critical
slowing down of the electronic mode is expected.

The likely absence of the slowing down in our data suggests that the
\citet{schaefer2014collective} model is not applicable in the present
case. The reason might be the neglecting of the CDW-coupled lattice-mode
damping in the model that might not be fulfilled in 2$H$-NbSe$_{2}$
due to the presence of the gapless $\Gamma$-pockets Fermi surface.

Focusing again to the rise-time dynamics we plot in Fig. \ref{fig:Selected--dependent-parameters.}
c) the $T$-dependence of the rise time at different excitation conditions.
The lowest-$F_{\mathrm{P}}$ 1.55 eV PPE rise time is found $T$ independent
($\tau_{\mathrm{r}}\sim1.7$ ps) up to $T\sim10$ K with a pronounced
drop with increasing $T$ beyond $\sim10$ K {[}see Fig. \ref{fig:Selected--dependent-parameters.}
c){]} to $\tau_{\mathrm{r}}\sim1$ ps at $T\sim15$ K dropping further
to $\tau_{\mathrm{r}}\sim0.8$~ps near $T_{\mathrm{CDW}}$. The rise
time drop with increasing $F_{\mathrm{P}}$ {[}see Fig. \ref{fig:Selected-F-dependent}
b){]} appears less pronounced at higher temperatures for the 1.55-PPE
case. The behavior is qualitatively consistent with the the non-equilibrium-phonons
dominated Rothwarf-Taylor pre-bottleneck dynamics\citep{kabanov2005kinetics}
discussed above.

There is a pronounced difference in the rise time behavior at 3.1
eV PPE, showing T-independent and slightly larger values\footnote{At comparable volume excitation densities.}
consistent with different rate-limiting $T$-independent and slower
high-energy quasiparticle relaxation pathways present at 3.1 eV PPE.

\section{Summary and Conclusions}

The coexisting CDW and SC phases in 2\emph{H}-NbSe$_{2}$ were investigated
by means of the narrow-band all-optical pump-probe spectroscopy extending
the parameter ranges of the previous similar works.\citep{anikin2020ultrafast,payne2020lattice}

Using 3.1 eV pump photon energy we reveal an unexpected high-energy
quasiparticle bottleneck due to a band gap in the unoccupied band
manifold which is present at all temperatures.

A systematic fluence dependence transient-reflectivity study shows
that the optical CDW suppression, with the absorbed energy density
threshold of $\sim1.2$ J/cm$^{3}$ (at $T=4$~K), is only weakly
non-thermal with a large amount of phonons excited concurrently. This
is different to the most of the common large-gap CDW materials and
superconductors. The low-fluence rise time dynamics data suggest that
the CDW suppression pathway is through hot phonons, similarly to the
conventional SCs\citep{demsar2003pairbreaking,beck2011energygap}.
The behavior is tentatively attributed to the presence of a relatively
large gapless Fermi-surface regions that enable efficient quasiparticle-energy
relaxation, without significant quasiparticle excitation in the CDW-gapped
hot spots.

To the best of our knowledge, we observe for the first time the concurrent
transient responses of the SC and CDW phases. However, only a very
weak signature of the SC state was observed, which did not allow for
a detailed SC state temporal dynamics analysis.

The CDW state relaxation is found to be dominated by the out-of plane
phonon diffusion processes. The heat transport is found to be much
less anisotropic than the charge transport, with the estimated low-temperature
out-of-plane thermal conductivity $\sim30$ times smaller than the
in-plane one.

The non-equilibrium-quasiparticle and the CDW order-parameter thermalization-timescale
slowing near $T\mathrm{_{CDW}}$\citep{schaefer2014collective,payne2020lattice}
appears unlikely, as most of the slowing down can be attributed to
the $T$ dependence of the diffusion constant.
\begin{acknowledgments}
The authors acknowledge the financial support of Slovenian Research
and Innovation Agency (research core funding No-P1-0040 and young
researcher funding No. PR-10496). We would also like to thank V. V.
Kabanov for fruitful discussions.
\end{acknowledgments}

\section{Saturation model\label{Appendix:sat-mod}}

Assuming laterally uniform beams and relatively narrow-band optical
pulses the transient reflectivity is given \citep{naseska2018ultrafast}
by:\begin{widetext}

\begin{eqnarray}
\frac{\Delta R}{R} & = & \frac{4\omega_{\mathrm{pr}}}{c_{0}\left|\mathcal{N}^{2}-1\right|}\int_{0}^{\infty}dze^{-\alpha z}\left[\Delta\epsilon_{\mathrm{r}}(z)\sin\left(2n\frac{\omega_{\mathrm{pr}}}{c_{0}}z-\beta\right)+\Delta\epsilon_{\mathrm{i}}(z)\cos\left(2n\frac{\omega_{\mathrm{pr}}}{c_{0}}z-\beta\right)\right],\label{eq:phaseFactors}\\
 &  & \mathcal{N}=n+i\kappa,\:\tan(\beta)=\frac{2n\kappa}{n^{2}-\kappa^{2}-1}.\nonumber 
\end{eqnarray}
\end{widetext}Here $n$ and $\kappa$ are the static refraction-index
real and imaginary parts at the probe PE, $\hbar\omega_{\mathrm{pr}}$,
respectively, $\alpha=2\kappa\frac{\omega_{\mathrm{pr}}}{c_{0}}$,
the probe extinction coefficient and $c_{0}$ the vacuum speed of
light. $\Delta\epsilon_{\mathrm{r}}(z)$ and $\Delta\epsilon_{\mathrm{i}}(z)$
correspond to the real and imaginary part of the photo-excitation-induced
dielectric function change, respectively.

Assuming, that $\Delta\epsilon_{\mathrm{r}}(z)$ and $\Delta\epsilon_{\mathrm{i}}(z)$
have the same $z$ dependence, $\Delta\epsilon(z)=\Delta\epsilon_{0}g(z)$,
Eq. (\ref{eq:phaseFactors}) is simplified to:, Eq. (\ref{eq:phaseFactors})
is simplified to:\begin{widetext} 
\begin{eqnarray}
\frac{\Delta R}{R} & = & \frac{4\omega_{\mathrm{pr}}\left|\Delta\epsilon_{0}\right|}{c_{0}\left|\mathcal{N}^{2}-1\right|}\int_{0}^{\infty}dze^{-\alpha z}\cos\left(2n\frac{\omega_{\mathrm{pr}}}{c_{0}}z-\phi\right)g(z),\label{eq:sat-dr}\\
\tan(\phi) & = & \frac{2n\kappa\Delta\epsilon_{0\mathrm{r}}-(n^{2}-\kappa^{2}-1)\Delta\epsilon_{0\mathrm{i}}}{2n\kappa\Delta\epsilon_{0\mathrm{i}}+(n^{2}-\kappa^{2}-1)\Delta\epsilon_{0\mathrm{r}}}.\label{eq:phi}
\end{eqnarray}
\end{widetext}In addition to the probe $n$ and $\alpha$ ($\kappa$),
which are given by the static optical properties, the integral kernel
depends on the phase shift $\phi$. The phase shift strongly influences
the kernel shape and, as a result, the depth sensitivity of the probe.
It cannot be determined from the static optical constants only and
needs to be determined from the transient data.

In the case of coaxial Gaussian beams with finite diameters (\ref{eq:sat-dr})
can be easily extended by an additional integration in the radial
direction\footnote{When both diameters are much larger than the corresponding wavelengths.}
where $g(r,z)$ is obtained from an appropriate effective model\citep{kusar2008controlled}
by taking into account the excitation fluence spatial dependence,
where $r$ corresponds to the radial distance from the beams center.

In the case of a linear excitation fluence response and a single excitation
P beam one can assume,
\begin{eqnarray}
g(r,z) & \propto & \Delta\epsilon(r,z)\propto U(r,z),\nonumber \\
U(r,z) & = & F_{\mathrm{P}}(1-R_{\mathrm{P}})\alpha_{\mathrm{P}}\exp\left[-\alpha_{\mathrm{P}}z-\nicefrac{2r^{2}}{\rho_{\mathrm{P}}^{2}}\right],\label{eq:lin_resp}
\end{eqnarray}
where $U(r,z)$ is the absorbed energy density. $R_{\mathrm{P}}$,
$\alpha_{\mathrm{P}}$ and $\rho_{\mathrm{P}}^{2}$ are the reflectivity,
extinction coefficient and diameter of the P beam, respectively.

To take into account suppression of the CDW resulting in a nonlinear
$\Delta\epsilon$ excitation dependence we assume a simple phenomenological
saturation model\citep{naseska2018ultrafast} where we approximate
the local amplitude of the transient dielectric function change, $\Delta\epsilon(r,z)$,
by a piece-wise linear function of the locally absorbed energy density,
$U(r,z)$, that has different slopes below and above $U_{\mathrm{th}}$:
\begin{eqnarray}
\Delta\epsilon(r,z) & = & \Delta\epsilon_{0}h(U(r,z)),\nonumber \\
h(u) & = & \begin{cases}
\frac{u}{U_{\mathrm{th}}}; & u<U_{\mathrm{th}}\\
1+a(\frac{u}{U_{\mathrm{th}}}-1); & u\geq U_{\mathrm{th}}
\end{cases}.\label{eq:sat}
\end{eqnarray}
Here $a$ corresponds to the relative slope in the normal state {[}see
insert to Fig. \ref{fig:Selected-F-dependent} b){]}.

For the case of the 3-pulse experiment one has to calculate the difference
(\ref{eq:DR3}) using (\ref{eq:sat-dr}) and taking (\ref{eq:sat})
with\footnote{Here we implicitly assume a chosen and fixed $t_{\mathrm{DP}}$.}
either $U(r,z)=U_{\mathrm{D}}(r,z)+U_{\mathrm{P}}(r,z)$ or $U(r,z)=U_{\mathrm{D}}(r,z)$
leading to:
\begin{eqnarray}
\Delta\epsilon_{3}(r,z) & = & \Delta\epsilon_{0}(h[U(r,z)+U_{\mathrm{P}}(r,z)]\nonumber \\
 &  & -h[U_{\mathrm{D}}(r,z)]).\label{eq:sat3p}
\end{eqnarray}
Here we assume that the probe arrives after P and D pulses and neglect
any temporal evolution.

For calculation of the diffusive recovery within the framework of
the saturation model we neglect the radial dependence and approximate
the depth profile with a Gaussian:
\begin{equation}
U(z,t)=U_{0}\exp\left[-\nicefrac{z^{2}}{z_{0}^{2}(1+\nicefrac{t}{\tau_{\mathrm{D}}})}\right]/\sqrt{1+\nicefrac{t}{\tau_{\mathrm{D}}}},\label{eq:diff}
\end{equation}
where, $z_{0}\sim\alpha^{-1},$ corresponds to the optical penetration
depth, $\tau_{\mathrm{D}}=\nicefrac{z_{0}^{2}}{4D}$, is given by
the diffusion constant $D$ and $U_{0}$ is the peak energy density.

In the case of an acoustic strain wave propagating perpendicular to
the surface\citep{thomsen1986picosecond} one can approximate $\Delta\epsilon(r,z)$
by a Heaviside function along $z$,

\begin{equation}
\Delta\epsilon(r,z)\propto\exp\left[-\nicefrac{2r^{2}}{\rho_{\mathrm{P}}^{2}}\right]\left[1-\mathrm{H}(z-c_{\mathrm{s}}t)\right],
\end{equation}
where $c_{\mathrm{s}}$ corresponds to the sound group velocity. The
oscillating part of the signal (\ref{eq:sat-dr}) is then,
\begin{equation}
\frac{\Delta R_{\mathrm{osc}}}{R}\propto e^{-\alpha c_{\mathrm{s}}t}\cos\left(\nicefrac{4\pi nc_{\mathrm{s}}t}{\lambda_{\mathrm{pr}}}-\phi\right),\label{eq:mov-mir}
\end{equation}
with $\lambda_{\mathrm{pr}}$ being the probe vacuum wavelength. The
sound speed is given by, 
\begin{equation}
c_{s}=\lambda_{pr}\nu_{s}/2n,\label{eq:sound-vel}
\end{equation}
 where $\nu_{s}$ corresponds to the measured coherent oscillation
frequency.

\bibliographystyle{apsrev4-1}
\bibliography{biblio}

\begin{thebibliography}{73}%
\makeatletter
\providecommand \@ifxundefined [1]{%
 \@ifx{#1\undefined}
}%
\providecommand \@ifnum [1]{%
 \ifnum #1\expandafter \@firstoftwo
 \else \expandafter \@secondoftwo
 \fi
}%
\providecommand \@ifx [1]{%
 \ifx #1\expandafter \@firstoftwo
 \else \expandafter \@secondoftwo
 \fi
}%
\providecommand \natexlab [1]{#1}%
\providecommand \enquote  [1]{``#1''}%
\providecommand \bibnamefont  [1]{#1}%
\providecommand \bibfnamefont [1]{#1}%
\providecommand \citenamefont [1]{#1}%
\providecommand \href@noop [0]{\@secondoftwo}%
\providecommand \href [0]{\begingroup \@sanitize@url \@href}%
\providecommand \@href[1]{\@@startlink{#1}\@@href}%
\providecommand \@@href[1]{\endgroup#1\@@endlink}%
\providecommand \@sanitize@url [0]{\catcode `\\12\catcode `\$12\catcode
  `\&12\catcode `\#12\catcode `\^12\catcode `\_12\catcode `\%12\relax}%
\providecommand \@@startlink[1]{}%
\providecommand \@@endlink[0]{}%
\providecommand \url  [0]{\begingroup\@sanitize@url \@url }%
\providecommand \@url [1]{\endgroup\@href {#1}{\urlprefix }}%
\providecommand \urlprefix  [0]{URL }%
\providecommand \Eprint [0]{\href }%
\providecommand \doibase [0]{http://dx.doi.org/}%
\providecommand \selectlanguage [0]{\@gobble}%
\providecommand \bibinfo  [0]{\@secondoftwo}%
\providecommand \bibfield  [0]{\@secondoftwo}%
\providecommand \translation [1]{[#1]}%
\providecommand \BibitemOpen [0]{}%
\providecommand \bibitemStop [0]{}%
\providecommand \bibitemNoStop [0]{.\EOS\space}%
\providecommand \EOS [0]{\spacefactor3000\relax}%
\providecommand \BibitemShut  [1]{\csname bibitem#1\endcsname}%
\let\auto@bib@innerbib\@empty
\bibitem [{\citenamefont {Berthier}\ \emph {et~al.}(1976)\citenamefont
  {Berthier}, \citenamefont {Molini{\'e}},\ and\ \citenamefont
  {J{\'e}rome}}]{berthier1976evidence}%
  \BibitemOpen
  \bibfield  {author} {\bibinfo {author} {\bibfnamefont {C.}~\bibnamefont
  {Berthier}}, \bibinfo {author} {\bibfnamefont {P.}~\bibnamefont
  {Molini{\'e}}}, \ and\ \bibinfo {author} {\bibfnamefont {D.}~\bibnamefont
  {J{\'e}rome}},\ }\href {\doibase 10.1016/0038-1098(76)90986-8} {\bibfield
  {journal} {\bibinfo  {journal} {Solid State Communications}\ }\textbf
  {\bibinfo {volume} {18}},\ \bibinfo {pages} {1393} (\bibinfo {year}
  {1976})}\BibitemShut {NoStop}%
\bibitem [{\citenamefont {Kusmartseva}\ \emph {et~al.}(2009)\citenamefont
  {Kusmartseva}, \citenamefont {Sipos}, \citenamefont {Berger}, \citenamefont
  {Forr{\'o}},\ and\ \citenamefont {Tuti{\v s}}}]{kusmartseva2009pressure}%
  \BibitemOpen
  \bibfield  {author} {\bibinfo {author} {\bibfnamefont {A.~F.}\ \bibnamefont
  {Kusmartseva}}, \bibinfo {author} {\bibfnamefont {B.}~\bibnamefont {Sipos}},
  \bibinfo {author} {\bibfnamefont {H.}~\bibnamefont {Berger}}, \bibinfo
  {author} {\bibfnamefont {L.}~\bibnamefont {Forr{\'o}}}, \ and\ \bibinfo
  {author} {\bibfnamefont {E.}~\bibnamefont {Tuti{\v s}}},\ }\href {\doibase
  10.1103/PhysRevLett.103.236401} {\bibfield  {journal} {\bibinfo  {journal}
  {Physical Review Letters}\ }\textbf {\bibinfo {volume} {103}},\ \bibinfo
  {pages} {236401} (\bibinfo {year} {2009})}\BibitemShut {NoStop}%
\bibitem [{\citenamefont {Sipos}\ \emph {et~al.}(2008)\citenamefont {Sipos},
  \citenamefont {Kusmartseva}, \citenamefont {Akrap}, \citenamefont {Berger},
  \citenamefont {Forr{\'o}},\ and\ \citenamefont {Tuti{\v
  s}}}]{sipos2008frommott}%
  \BibitemOpen
  \bibfield  {author} {\bibinfo {author} {\bibfnamefont {B.}~\bibnamefont
  {Sipos}}, \bibinfo {author} {\bibfnamefont {A.~F.}\ \bibnamefont
  {Kusmartseva}}, \bibinfo {author} {\bibfnamefont {A.}~\bibnamefont {Akrap}},
  \bibinfo {author} {\bibfnamefont {H.}~\bibnamefont {Berger}}, \bibinfo
  {author} {\bibfnamefont {L.}~\bibnamefont {Forr{\'o}}}, \ and\ \bibinfo
  {author} {\bibfnamefont {E.}~\bibnamefont {Tuti{\v s}}},\ }\href {\doibase
  10.1038/nmat2318} {\bibfield  {journal} {\bibinfo  {journal} {Nature
  Materials}\ }\textbf {\bibinfo {volume} {7}},\ \bibinfo {pages} {960}
  (\bibinfo {year} {2008})}\BibitemShut {NoStop}%
\bibitem [{\citenamefont {Gao}\ \emph {et~al.}(2018)\citenamefont {Gao},
  \citenamefont {Flicker}, \citenamefont {Sankar}, \citenamefont {Zhao},
  \citenamefont {Ren}, \citenamefont {Rachmilowitz}, \citenamefont
  {Balachandar}, \citenamefont {Chou}, \citenamefont {Burch}, \citenamefont
  {Wang}, \citenamefont {{van Wezel}},\ and\ \citenamefont
  {Zeljkovic}}]{gao2018atomicscale}%
  \BibitemOpen
  \bibfield  {author} {\bibinfo {author} {\bibfnamefont {S.}~\bibnamefont
  {Gao}}, \bibinfo {author} {\bibfnamefont {F.}~\bibnamefont {Flicker}},
  \bibinfo {author} {\bibfnamefont {R.}~\bibnamefont {Sankar}}, \bibinfo
  {author} {\bibfnamefont {H.}~\bibnamefont {Zhao}}, \bibinfo {author}
  {\bibfnamefont {Z.}~\bibnamefont {Ren}}, \bibinfo {author} {\bibfnamefont
  {B.}~\bibnamefont {Rachmilowitz}}, \bibinfo {author} {\bibfnamefont
  {S.}~\bibnamefont {Balachandar}}, \bibinfo {author} {\bibfnamefont
  {F.}~\bibnamefont {Chou}}, \bibinfo {author} {\bibfnamefont {K.~S.}\
  \bibnamefont {Burch}}, \bibinfo {author} {\bibfnamefont {Z.}~\bibnamefont
  {Wang}}, \bibinfo {author} {\bibfnamefont {J.}~\bibnamefont {{van Wezel}}}, \
  and\ \bibinfo {author} {\bibfnamefont {I.}~\bibnamefont {Zeljkovic}},\ }\href
  {\doibase 10.1073/pnas.1718931115} {\bibfield  {journal} {\bibinfo  {journal}
  {Proceedings of the National Academy of Sciences}\ }\textbf {\bibinfo
  {volume} {115}},\ \bibinfo {pages} {6986} (\bibinfo {year}
  {2018})}\BibitemShut {NoStop}%
\bibitem [{\citenamefont {Qian}\ \emph {et~al.}(2021)\citenamefont {Qian},
  \citenamefont {Christensen}, \citenamefont {Hu}, \citenamefont {Saha},
  \citenamefont {Andersen}, \citenamefont {Fernandes}, \citenamefont {Birol},\
  and\ \citenamefont {Ni}}]{qian2021revealing}%
  \BibitemOpen
  \bibfield  {author} {\bibinfo {author} {\bibfnamefont {T.}~\bibnamefont
  {Qian}}, \bibinfo {author} {\bibfnamefont {M.~H.}\ \bibnamefont
  {Christensen}}, \bibinfo {author} {\bibfnamefont {C.}~\bibnamefont {Hu}},
  \bibinfo {author} {\bibfnamefont {A.}~\bibnamefont {Saha}}, \bibinfo {author}
  {\bibfnamefont {B.~M.}\ \bibnamefont {Andersen}}, \bibinfo {author}
  {\bibfnamefont {R.~M.}\ \bibnamefont {Fernandes}}, \bibinfo {author}
  {\bibfnamefont {T.}~\bibnamefont {Birol}}, \ and\ \bibinfo {author}
  {\bibfnamefont {N.}~\bibnamefont {Ni}},\ }\href {\doibase
  10.1103/PhysRevB.104.144506} {\bibfield  {journal} {\bibinfo  {journal}
  {Physical Review B}\ }\textbf {\bibinfo {volume} {104}},\ \bibinfo {pages}
  {144506} (\bibinfo {year} {2021})}\BibitemShut {NoStop}%
\bibitem [{\citenamefont {Qiao}\ \emph {et~al.}(2017)\citenamefont {Qiao},
  \citenamefont {Li}, \citenamefont {Wang}, \citenamefont {Ruan}, \citenamefont
  {Ye}, \citenamefont {Cai}, \citenamefont {Hao}, \citenamefont {Yao},
  \citenamefont {Chen}, \citenamefont {Wu}, \citenamefont {Wang},\ and\
  \citenamefont {Liu}}]{qiao2017mottness}%
  \BibitemOpen
  \bibfield  {author} {\bibinfo {author} {\bibfnamefont {S.}~\bibnamefont
  {Qiao}}, \bibinfo {author} {\bibfnamefont {X.}~\bibnamefont {Li}}, \bibinfo
  {author} {\bibfnamefont {N.}~\bibnamefont {Wang}}, \bibinfo {author}
  {\bibfnamefont {W.}~\bibnamefont {Ruan}}, \bibinfo {author} {\bibfnamefont
  {C.}~\bibnamefont {Ye}}, \bibinfo {author} {\bibfnamefont {P.}~\bibnamefont
  {Cai}}, \bibinfo {author} {\bibfnamefont {Z.}~\bibnamefont {Hao}}, \bibinfo
  {author} {\bibfnamefont {H.}~\bibnamefont {Yao}}, \bibinfo {author}
  {\bibfnamefont {X.}~\bibnamefont {Chen}}, \bibinfo {author} {\bibfnamefont
  {J.}~\bibnamefont {Wu}}, \bibinfo {author} {\bibfnamefont {Y.}~\bibnamefont
  {Wang}}, \ and\ \bibinfo {author} {\bibfnamefont {Z.}~\bibnamefont {Liu}},\
  }\href {\doibase 10.1103/PhysRevX.7.041054} {\bibfield  {journal} {\bibinfo
  {journal} {Physical Review X}\ }\textbf {\bibinfo {volume} {7}} (\bibinfo
  {year} {2017}),\ 10.1103/PhysRevX.7.041054}\BibitemShut {NoStop}%
\bibitem [{\citenamefont {Yu}\ \emph {et~al.}(2015)\citenamefont {Yu},
  \citenamefont {Yang}, \citenamefont {Lu}, \citenamefont {Yan}, \citenamefont
  {Cho}, \citenamefont {Ma}, \citenamefont {Niu}, \citenamefont {Kim},
  \citenamefont {Son}, \citenamefont {Feng}, \citenamefont {Li}, \citenamefont
  {Cheong}, \citenamefont {Chen},\ and\ \citenamefont
  {Zhang}}]{yu2015gatetunable}%
  \BibitemOpen
  \bibfield  {author} {\bibinfo {author} {\bibfnamefont {Y.}~\bibnamefont
  {Yu}}, \bibinfo {author} {\bibfnamefont {F.}~\bibnamefont {Yang}}, \bibinfo
  {author} {\bibfnamefont {X.~F.}\ \bibnamefont {Lu}}, \bibinfo {author}
  {\bibfnamefont {Y.~J.}\ \bibnamefont {Yan}}, \bibinfo {author} {\bibfnamefont
  {Y.-H.}\ \bibnamefont {Cho}}, \bibinfo {author} {\bibfnamefont
  {L.}~\bibnamefont {Ma}}, \bibinfo {author} {\bibfnamefont {X.}~\bibnamefont
  {Niu}}, \bibinfo {author} {\bibfnamefont {S.}~\bibnamefont {Kim}}, \bibinfo
  {author} {\bibfnamefont {Y.-W.}\ \bibnamefont {Son}}, \bibinfo {author}
  {\bibfnamefont {D.}~\bibnamefont {Feng}}, \bibinfo {author} {\bibfnamefont
  {S.}~\bibnamefont {Li}}, \bibinfo {author} {\bibfnamefont {S.-W.}\
  \bibnamefont {Cheong}}, \bibinfo {author} {\bibfnamefont {X.~H.}\
  \bibnamefont {Chen}}, \ and\ \bibinfo {author} {\bibfnamefont
  {Y.}~\bibnamefont {Zhang}},\ }\href {\doibase 10.1038/nnano.2014.323}
  {\bibfield  {journal} {\bibinfo  {journal} {Nature Nanotechnology}\ }\textbf
  {\bibinfo {volume} {10}},\ \bibinfo {pages} {270} (\bibinfo {year}
  {2015})}\BibitemShut {NoStop}%
\bibitem [{\citenamefont {Chatterjee}\ \emph {et~al.}(2015)\citenamefont
  {Chatterjee}, \citenamefont {Zhao}, \citenamefont {Iavarone}, \citenamefont
  {Di~Capua}, \citenamefont {Castellan}, \citenamefont {Karapetrov},
  \citenamefont {Malliakas}, \citenamefont {Kanatzidis}, \citenamefont {Claus},
  \citenamefont {Ruff}, \citenamefont {Weber}, \citenamefont {{van Wezel}},
  \citenamefont {Campuzano}, \citenamefont {Osborn}, \citenamefont {Randeria},
  \citenamefont {Trivedi}, \citenamefont {Norman},\ and\ \citenamefont
  {Rosenkranz}}]{chatterjee2015emergence}%
  \BibitemOpen
  \bibfield  {author} {\bibinfo {author} {\bibfnamefont {U.}~\bibnamefont
  {Chatterjee}}, \bibinfo {author} {\bibfnamefont {J.}~\bibnamefont {Zhao}},
  \bibinfo {author} {\bibfnamefont {M.}~\bibnamefont {Iavarone}}, \bibinfo
  {author} {\bibfnamefont {R.}~\bibnamefont {Di~Capua}}, \bibinfo {author}
  {\bibfnamefont {J.~P.}\ \bibnamefont {Castellan}}, \bibinfo {author}
  {\bibfnamefont {G.}~\bibnamefont {Karapetrov}}, \bibinfo {author}
  {\bibfnamefont {C.~D.}\ \bibnamefont {Malliakas}}, \bibinfo {author}
  {\bibfnamefont {M.~G.}\ \bibnamefont {Kanatzidis}}, \bibinfo {author}
  {\bibfnamefont {H.}~\bibnamefont {Claus}}, \bibinfo {author} {\bibfnamefont
  {J.~P.~C.}\ \bibnamefont {Ruff}}, \bibinfo {author} {\bibfnamefont
  {F.}~\bibnamefont {Weber}}, \bibinfo {author} {\bibfnamefont
  {J.}~\bibnamefont {{van Wezel}}}, \bibinfo {author} {\bibfnamefont {J.~C.}\
  \bibnamefont {Campuzano}}, \bibinfo {author} {\bibfnamefont {R.}~\bibnamefont
  {Osborn}}, \bibinfo {author} {\bibfnamefont {M.}~\bibnamefont {Randeria}},
  \bibinfo {author} {\bibfnamefont {N.}~\bibnamefont {Trivedi}}, \bibinfo
  {author} {\bibfnamefont {M.~R.}\ \bibnamefont {Norman}}, \ and\ \bibinfo
  {author} {\bibfnamefont {S.}~\bibnamefont {Rosenkranz}},\ }\href {\doibase
  10.1038/ncomms7313} {\bibfield  {journal} {\bibinfo  {journal} {Nature
  Communications}\ }\textbf {\bibinfo {volume} {6}},\ \bibinfo {pages} {6313}
  (\bibinfo {year} {2015})}\BibitemShut {NoStop}%
\bibitem [{\citenamefont {Meyer}\ \emph {et~al.}(1975)\citenamefont {Meyer},
  \citenamefont {Howard}, \citenamefont {Stewart}, \citenamefont {Acrivos},\
  and\ \citenamefont {Geballe}}]{meyer1975properties}%
  \BibitemOpen
  \bibfield  {author} {\bibinfo {author} {\bibfnamefont {S.~F.}\ \bibnamefont
  {Meyer}}, \bibinfo {author} {\bibfnamefont {R.~E.}\ \bibnamefont {Howard}},
  \bibinfo {author} {\bibfnamefont {G.~R.}\ \bibnamefont {Stewart}}, \bibinfo
  {author} {\bibfnamefont {J.~V.}\ \bibnamefont {Acrivos}}, \ and\ \bibinfo
  {author} {\bibfnamefont {T.~H.}\ \bibnamefont {Geballe}},\ }\href {\doibase
  10.1063/1.430342} {\bibfield  {journal} {\bibinfo  {journal} {The Journal of
  Chemical Physics}\ }\textbf {\bibinfo {volume} {62}},\ \bibinfo {pages}
  {4411} (\bibinfo {year} {1975})}\BibitemShut {NoStop}%
\bibitem [{\citenamefont {Morosan}\ \emph {et~al.}(2006)\citenamefont
  {Morosan}, \citenamefont {Zandbergen}, \citenamefont {Dennis}, \citenamefont
  {Bos}, \citenamefont {Onose}, \citenamefont {Klimczuk}, \citenamefont
  {Ramirez}, \citenamefont {Ong},\ and\ \citenamefont
  {Cava}}]{morosan2006superconductivity}%
  \BibitemOpen
  \bibfield  {author} {\bibinfo {author} {\bibfnamefont {E.}~\bibnamefont
  {Morosan}}, \bibinfo {author} {\bibfnamefont {H.~W.}\ \bibnamefont
  {Zandbergen}}, \bibinfo {author} {\bibfnamefont {B.~S.}\ \bibnamefont
  {Dennis}}, \bibinfo {author} {\bibfnamefont {J.~W.~G.}\ \bibnamefont {Bos}},
  \bibinfo {author} {\bibfnamefont {Y.}~\bibnamefont {Onose}}, \bibinfo
  {author} {\bibfnamefont {T.}~\bibnamefont {Klimczuk}}, \bibinfo {author}
  {\bibfnamefont {A.~P.}\ \bibnamefont {Ramirez}}, \bibinfo {author}
  {\bibfnamefont {N.~P.}\ \bibnamefont {Ong}}, \ and\ \bibinfo {author}
  {\bibfnamefont {R.~J.}\ \bibnamefont {Cava}},\ }\href {\doibase
  10.1038/nphys360} {\bibfield  {journal} {\bibinfo  {journal} {Nature
  Physics}\ }\textbf {\bibinfo {volume} {2}},\ \bibinfo {pages} {544} (\bibinfo
  {year} {2006})}\BibitemShut {NoStop}%
\bibitem [{\citenamefont {Stojchevska}\ \emph {et~al.}(2014)\citenamefont
  {Stojchevska}, \citenamefont {Vaskivskyi}, \citenamefont {Mertelj},
  \citenamefont {Kusar}, \citenamefont {Svetin}, \citenamefont {Brazovskii},\
  and\ \citenamefont {Mihailovic}}]{stojchevska2014ultrafast}%
  \BibitemOpen
  \bibfield  {author} {\bibinfo {author} {\bibfnamefont {L.}~\bibnamefont
  {Stojchevska}}, \bibinfo {author} {\bibfnamefont {I.}~\bibnamefont
  {Vaskivskyi}}, \bibinfo {author} {\bibfnamefont {T.}~\bibnamefont {Mertelj}},
  \bibinfo {author} {\bibfnamefont {P.}~\bibnamefont {Kusar}}, \bibinfo
  {author} {\bibfnamefont {D.}~\bibnamefont {Svetin}}, \bibinfo {author}
  {\bibfnamefont {S.}~\bibnamefont {Brazovskii}}, \ and\ \bibinfo {author}
  {\bibfnamefont {D.}~\bibnamefont {Mihailovic}},\ }\href {\doibase
  10.1126/science.1241591} {\bibfield  {journal} {\bibinfo  {journal} {Science
  (New York, N.Y.)}\ }\textbf {\bibinfo {volume} {344}},\ \bibinfo {pages}
  {177} (\bibinfo {year} {2014})}\BibitemShut {NoStop}%
\bibitem [{\citenamefont {Mraz}\ \emph {et~al.}(2022)\citenamefont {Mraz},
  \citenamefont {Venturini}, \citenamefont {Svetin}, \citenamefont {Sever},
  \citenamefont {Mihailovic}, \citenamefont {Vaskivskyi}, \citenamefont
  {Ambrozic}, \citenamefont {Dra{\v z}i{\'c}}, \citenamefont {D'Antuono},
  \citenamefont {Stornaiuolo}, \citenamefont {Tafuri}, \citenamefont {Kazazis},
  \citenamefont {Ravnik}, \citenamefont {Ekinci},\ and\ \citenamefont
  {Mihailovic}}]{mraz2022chargeconfiguration}%
  \BibitemOpen
  \bibfield  {author} {\bibinfo {author} {\bibfnamefont {A.}~\bibnamefont
  {Mraz}}, \bibinfo {author} {\bibfnamefont {R.}~\bibnamefont {Venturini}},
  \bibinfo {author} {\bibfnamefont {D.}~\bibnamefont {Svetin}}, \bibinfo
  {author} {\bibfnamefont {V.}~\bibnamefont {Sever}}, \bibinfo {author}
  {\bibfnamefont {I.~A.}\ \bibnamefont {Mihailovic}}, \bibinfo {author}
  {\bibfnamefont {I.}~\bibnamefont {Vaskivskyi}}, \bibinfo {author}
  {\bibfnamefont {B.}~\bibnamefont {Ambrozic}}, \bibinfo {author}
  {\bibfnamefont {G.}~\bibnamefont {Dra{\v z}i{\'c}}}, \bibinfo {author}
  {\bibfnamefont {M.}~\bibnamefont {D'Antuono}}, \bibinfo {author}
  {\bibfnamefont {D.}~\bibnamefont {Stornaiuolo}}, \bibinfo {author}
  {\bibfnamefont {F.}~\bibnamefont {Tafuri}}, \bibinfo {author} {\bibfnamefont
  {D.}~\bibnamefont {Kazazis}}, \bibinfo {author} {\bibfnamefont
  {J.}~\bibnamefont {Ravnik}}, \bibinfo {author} {\bibfnamefont
  {Y.}~\bibnamefont {Ekinci}}, \ and\ \bibinfo {author} {\bibfnamefont
  {D.}~\bibnamefont {Mihailovic}},\ }\href {\doibase
  10.1021/acs.nanolett.2c01116} {\bibfield  {journal} {\bibinfo  {journal}
  {Nano Letters}\ }\textbf {\bibinfo {volume} {22}},\ \bibinfo {pages} {4814}
  (\bibinfo {year} {2022})}\BibitemShut {NoStop}%
\bibitem [{\citenamefont {Vaskivskyi}\ \emph {et~al.}(2016)\citenamefont
  {Vaskivskyi}, \citenamefont {Mihailovic}, \citenamefont {Brazovskii},
  \citenamefont {Gospodaric}, \citenamefont {Mertelj}, \citenamefont {Svetin},
  \citenamefont {Sutar},\ and\ \citenamefont
  {Mihailovic}}]{vaskivskyi2016fastelectronic}%
  \BibitemOpen
  \bibfield  {author} {\bibinfo {author} {\bibfnamefont {I.}~\bibnamefont
  {Vaskivskyi}}, \bibinfo {author} {\bibfnamefont {I.~A.}\ \bibnamefont
  {Mihailovic}}, \bibinfo {author} {\bibfnamefont {S.}~\bibnamefont
  {Brazovskii}}, \bibinfo {author} {\bibfnamefont {J.}~\bibnamefont
  {Gospodaric}}, \bibinfo {author} {\bibfnamefont {T.}~\bibnamefont {Mertelj}},
  \bibinfo {author} {\bibfnamefont {D.}~\bibnamefont {Svetin}}, \bibinfo
  {author} {\bibfnamefont {P.}~\bibnamefont {Sutar}}, \ and\ \bibinfo {author}
  {\bibfnamefont {D.}~\bibnamefont {Mihailovic}},\ }\href {\doibase
  10.1038/ncomms11442} {\bibfield  {journal} {\bibinfo  {journal} {Nature
  Communications}\ }\textbf {\bibinfo {volume} {7}},\ \bibinfo {pages} {11442}
  (\bibinfo {year} {2016})}\BibitemShut {NoStop}%
\bibitem [{\citenamefont {Yoshida}\ \emph {et~al.}(2015)\citenamefont
  {Yoshida}, \citenamefont {Suzuki}, \citenamefont {Zhang}, \citenamefont
  {Nakano},\ and\ \citenamefont {Iwasa}}]{yoshida2015memristive}%
  \BibitemOpen
  \bibfield  {author} {\bibinfo {author} {\bibfnamefont {M.}~\bibnamefont
  {Yoshida}}, \bibinfo {author} {\bibfnamefont {R.}~\bibnamefont {Suzuki}},
  \bibinfo {author} {\bibfnamefont {Y.}~\bibnamefont {Zhang}}, \bibinfo
  {author} {\bibfnamefont {M.}~\bibnamefont {Nakano}}, \ and\ \bibinfo {author}
  {\bibfnamefont {Y.}~\bibnamefont {Iwasa}},\ }\href {\doibase
  10.1126/sciadv.1500606} {\bibfield  {journal} {\bibinfo  {journal} {Science
  Advances}\ }\textbf {\bibinfo {volume} {1}},\ \bibinfo {pages} {e1500606}
  (\bibinfo {year} {2015})}\BibitemShut {NoStop}%
\bibitem [{\citenamefont {Moncton}\ \emph {et~al.}(1975)\citenamefont
  {Moncton}, \citenamefont {Axe},\ and\ \citenamefont
  {DiSalvo}}]{moncton1975studyof}%
  \BibitemOpen
  \bibfield  {author} {\bibinfo {author} {\bibfnamefont {D.~E.}\ \bibnamefont
  {Moncton}}, \bibinfo {author} {\bibfnamefont {J.~D.}\ \bibnamefont {Axe}}, \
  and\ \bibinfo {author} {\bibfnamefont {F.~J.}\ \bibnamefont {DiSalvo}},\
  }\href {\doibase 10.1103/PhysRevLett.34.734} {\bibfield  {journal} {\bibinfo
  {journal} {Physical Review Letters}\ }\textbf {\bibinfo {volume} {34}},\
  \bibinfo {pages} {734} (\bibinfo {year} {1975})}\BibitemShut {NoStop}%
\bibitem [{\citenamefont {Rahn}\ \emph {et~al.}(2012)\citenamefont {Rahn},
  \citenamefont {Hellmann}, \citenamefont {Kall{\"a}ne}, \citenamefont {Sohrt},
  \citenamefont {Kim}, \citenamefont {Kipp},\ and\ \citenamefont
  {Rossnagel}}]{rahn2012gapsand}%
  \BibitemOpen
  \bibfield  {author} {\bibinfo {author} {\bibfnamefont {D.~J.}\ \bibnamefont
  {Rahn}}, \bibinfo {author} {\bibfnamefont {S.}~\bibnamefont {Hellmann}},
  \bibinfo {author} {\bibfnamefont {M.}~\bibnamefont {Kall{\"a}ne}}, \bibinfo
  {author} {\bibfnamefont {C.}~\bibnamefont {Sohrt}}, \bibinfo {author}
  {\bibfnamefont {T.~K.}\ \bibnamefont {Kim}}, \bibinfo {author} {\bibfnamefont
  {L.}~\bibnamefont {Kipp}}, \ and\ \bibinfo {author} {\bibfnamefont
  {K.}~\bibnamefont {Rossnagel}},\ }\href {\doibase 10.1103/PhysRevB.85.224532}
  {\bibfield  {journal} {\bibinfo  {journal} {Physical Review B}\ }\textbf
  {\bibinfo {volume} {85}},\ \bibinfo {pages} {224532} (\bibinfo {year}
  {2012})}\BibitemShut {NoStop}%
\bibitem [{\citenamefont {Mogami}\ \emph {et~al.}(2021)\citenamefont {Mogami},
  \citenamefont {Ohta},\ and\ \citenamefont {Sakata}}]{mogami2021appearance}%
  \BibitemOpen
  \bibfield  {author} {\bibinfo {author} {\bibfnamefont {K.}~\bibnamefont
  {Mogami}}, \bibinfo {author} {\bibfnamefont {S.}~\bibnamefont {Ohta}}, \ and\
  \bibinfo {author} {\bibfnamefont {H.}~\bibnamefont {Sakata}},\ }\href
  {\doibase 10.1016/j.susc.2020.121636} {\bibfield  {journal} {\bibinfo
  {journal} {Surface Science}\ }\textbf {\bibinfo {volume} {707}},\ \bibinfo
  {pages} {121636} (\bibinfo {year} {2021})}\BibitemShut {NoStop}%
\bibitem [{\citenamefont {Bischoff}\ \emph {et~al.}(2017)\citenamefont
  {Bischoff}, \citenamefont {Auw{\"a}rter}, \citenamefont {Barth},
  \citenamefont {Schiffrin}, \citenamefont {Fuhrer},\ and\ \citenamefont
  {Weber}}]{bischoff2017nanoscale}%
  \BibitemOpen
  \bibfield  {author} {\bibinfo {author} {\bibfnamefont {F.}~\bibnamefont
  {Bischoff}}, \bibinfo {author} {\bibfnamefont {W.}~\bibnamefont
  {Auw{\"a}rter}}, \bibinfo {author} {\bibfnamefont {J.~V.}\ \bibnamefont
  {Barth}}, \bibinfo {author} {\bibfnamefont {A.}~\bibnamefont {Schiffrin}},
  \bibinfo {author} {\bibfnamefont {M.}~\bibnamefont {Fuhrer}}, \ and\ \bibinfo
  {author} {\bibfnamefont {B.}~\bibnamefont {Weber}},\ }\href {\doibase
  10.1021/acs.chemmater.7b03061} {\bibfield  {journal} {\bibinfo  {journal}
  {Chemistry of Materials}\ }\textbf {\bibinfo {volume} {29}},\ \bibinfo
  {pages} {9907} (\bibinfo {year} {2017})}\BibitemShut {NoStop}%
\bibitem [{\citenamefont {Liu}\ \emph {et~al.}(2013)\citenamefont {Liu},
  \citenamefont {Ang}, \citenamefont {Lu}, \citenamefont {Song}, \citenamefont
  {Li},\ and\ \citenamefont {Sun}}]{liu2013superconductivity}%
  \BibitemOpen
  \bibfield  {author} {\bibinfo {author} {\bibfnamefont {Y.}~\bibnamefont
  {Liu}}, \bibinfo {author} {\bibfnamefont {R.}~\bibnamefont {Ang}}, \bibinfo
  {author} {\bibfnamefont {W.~J.}\ \bibnamefont {Lu}}, \bibinfo {author}
  {\bibfnamefont {W.~H.}\ \bibnamefont {Song}}, \bibinfo {author}
  {\bibfnamefont {L.~J.}\ \bibnamefont {Li}}, \ and\ \bibinfo {author}
  {\bibfnamefont {Y.~P.}\ \bibnamefont {Sun}},\ }\href {\doibase
  10.1063/1.4805003} {\bibfield  {journal} {\bibinfo  {journal} {Applied
  Physics Letters}\ }\textbf {\bibinfo {volume} {102}},\ \bibinfo {pages}
  {192602} (\bibinfo {year} {2013})}\BibitemShut {NoStop}%
\bibitem [{\citenamefont {Littlewood}\ and\ \citenamefont
  {Varma}(1982)}]{littlewood1982amplitude}%
  \BibitemOpen
  \bibfield  {author} {\bibinfo {author} {\bibfnamefont {P.~B.}\ \bibnamefont
  {Littlewood}}\ and\ \bibinfo {author} {\bibfnamefont {C.~M.}\ \bibnamefont
  {Varma}},\ }\href {\doibase 10.1103/PhysRevB.26.4883} {\bibfield  {journal}
  {\bibinfo  {journal} {Physical Review B}\ }\textbf {\bibinfo {volume} {26}},\
  \bibinfo {pages} {4883} (\bibinfo {year} {1982})}\BibitemShut {NoStop}%
\bibitem [{\citenamefont {Pekker}\ and\ \citenamefont
  {Varma}(2015)}]{pekker2015amplitudehiggs}%
  \BibitemOpen
  \bibfield  {author} {\bibinfo {author} {\bibfnamefont {D.}~\bibnamefont
  {Pekker}}\ and\ \bibinfo {author} {\bibfnamefont {C.}~\bibnamefont {Varma}},\
  }\href {\doibase 10.1146/annurev-conmatphys-031214-014350} {\bibfield
  {journal} {\bibinfo  {journal} {Annual Review of Condensed Matter Physics}\
  }\textbf {\bibinfo {volume} {6}},\ \bibinfo {pages} {269} (\bibinfo {year}
  {2015})}\BibitemShut {NoStop}%
\bibitem [{\citenamefont {Grasset}\ \emph {et~al.}(2018)\citenamefont
  {Grasset}, \citenamefont {Cea}, \citenamefont {Gallais}, \citenamefont
  {Cazayous}, \citenamefont {Sacuto}, \citenamefont {Cario}, \citenamefont
  {Benfatto},\ and\ \citenamefont {M{\'e}asson}}]{grasset2018higgsmode}%
  \BibitemOpen
  \bibfield  {author} {\bibinfo {author} {\bibfnamefont {R.}~\bibnamefont
  {Grasset}}, \bibinfo {author} {\bibfnamefont {T.}~\bibnamefont {Cea}},
  \bibinfo {author} {\bibfnamefont {Y.}~\bibnamefont {Gallais}}, \bibinfo
  {author} {\bibfnamefont {M.}~\bibnamefont {Cazayous}}, \bibinfo {author}
  {\bibfnamefont {A.}~\bibnamefont {Sacuto}}, \bibinfo {author} {\bibfnamefont
  {L.}~\bibnamefont {Cario}}, \bibinfo {author} {\bibfnamefont
  {L.}~\bibnamefont {Benfatto}}, \ and\ \bibinfo {author} {\bibfnamefont
  {M.-A.}\ \bibnamefont {M{\'e}asson}},\ }\href {\doibase
  10.1103/PhysRevB.97.094502} {\bibfield  {journal} {\bibinfo  {journal}
  {Physical Review B}\ }\textbf {\bibinfo {volume} {97}},\ \bibinfo {pages}
  {094502} (\bibinfo {year} {2018})}\BibitemShut {NoStop}%
\bibitem [{\citenamefont {M{\'e}asson}\ \emph {et~al.}(2014)\citenamefont
  {M{\'e}asson}, \citenamefont {Gallais}, \citenamefont {Cazayous},
  \citenamefont {Clair}, \citenamefont {Rodi{\`e}re}, \citenamefont {Cario},\
  and\ \citenamefont {Sacuto}}]{measson2014amplitude}%
  \BibitemOpen
  \bibfield  {author} {\bibinfo {author} {\bibfnamefont {M.-A.}\ \bibnamefont
  {M{\'e}asson}}, \bibinfo {author} {\bibfnamefont {Y.}~\bibnamefont
  {Gallais}}, \bibinfo {author} {\bibfnamefont {M.}~\bibnamefont {Cazayous}},
  \bibinfo {author} {\bibfnamefont {B.}~\bibnamefont {Clair}}, \bibinfo
  {author} {\bibfnamefont {P.}~\bibnamefont {Rodi{\`e}re}}, \bibinfo {author}
  {\bibfnamefont {L.}~\bibnamefont {Cario}}, \ and\ \bibinfo {author}
  {\bibfnamefont {A.}~\bibnamefont {Sacuto}},\ }\href {\doibase
  10.1103/PhysRevB.89.060503} {\bibfield  {journal} {\bibinfo  {journal}
  {Physical Review B}\ }\textbf {\bibinfo {volume} {89}},\ \bibinfo {pages}
  {060503} (\bibinfo {year} {2014})}\BibitemShut {NoStop}%
\bibitem [{\citenamefont {Sooryakumar}\ and\ \citenamefont
  {Klein}(1980)}]{sooryakumar1980ramanscattering}%
  \BibitemOpen
  \bibfield  {author} {\bibinfo {author} {\bibfnamefont {R.}~\bibnamefont
  {Sooryakumar}}\ and\ \bibinfo {author} {\bibfnamefont {M.~V.}\ \bibnamefont
  {Klein}},\ }\href {\doibase 10.1103/PhysRevLett.45.660} {\bibfield  {journal}
  {\bibinfo  {journal} {Physical Review Letters}\ }\textbf {\bibinfo {volume}
  {45}},\ \bibinfo {pages} {660} (\bibinfo {year} {1980})}\BibitemShut
  {NoStop}%
\bibitem [{\citenamefont {Anikin}\ \emph {et~al.}(2020)\citenamefont {Anikin},
  \citenamefont {Schaller}, \citenamefont {Wiederrecht}, \citenamefont
  {Margine}, \citenamefont {Mazin},\ and\ \citenamefont
  {Karapetrov}}]{anikin2020ultrafast}%
  \BibitemOpen
  \bibfield  {author} {\bibinfo {author} {\bibfnamefont {A.}~\bibnamefont
  {Anikin}}, \bibinfo {author} {\bibfnamefont {R.~D.}\ \bibnamefont
  {Schaller}}, \bibinfo {author} {\bibfnamefont {G.~P.}\ \bibnamefont
  {Wiederrecht}}, \bibinfo {author} {\bibfnamefont {E.~R.}\ \bibnamefont
  {Margine}}, \bibinfo {author} {\bibfnamefont {I.~I.}\ \bibnamefont {Mazin}},
  \ and\ \bibinfo {author} {\bibfnamefont {G.}~\bibnamefont {Karapetrov}},\
  }\href {\doibase 10.1103/PhysRevB.102.205139} {\bibfield  {journal} {\bibinfo
   {journal} {Physical Review B}\ }\textbf {\bibinfo {volume} {102}},\ \bibinfo
  {pages} {205139} (\bibinfo {year} {2020})}\BibitemShut {NoStop}%
\bibitem [{\citenamefont {Payne}\ \emph {et~al.}(2020)\citenamefont {Payne},
  \citenamefont {Barone}, \citenamefont {Benfatto}, \citenamefont
  {Parmigiani},\ and\ \citenamefont {Cilento}}]{payne2020lattice}%
  \BibitemOpen
  \bibfield  {author} {\bibinfo {author} {\bibfnamefont {D.~T.}\ \bibnamefont
  {Payne}}, \bibinfo {author} {\bibfnamefont {P.}~\bibnamefont {Barone}},
  \bibinfo {author} {\bibfnamefont {L.}~\bibnamefont {Benfatto}}, \bibinfo
  {author} {\bibfnamefont {F.}~\bibnamefont {Parmigiani}}, \ and\ \bibinfo
  {author} {\bibfnamefont {F.}~\bibnamefont {Cilento}},\ }\href
  {http://arxiv.org/abs/2010.09826} {\enquote {\bibinfo {title} {Lattice
  contribution to the unconventional charge density wave transition in
  \${{2H}}\$-{{NbSe}}\$\_2\$: A non-equilibrium optical approach},}\ }
  (\bibinfo {year} {2020}),\ \Eprint {http://arxiv.org/abs/2010.09826}
  {arXiv:2010.09826 [cond-mat]} \BibitemShut {NoStop}%
\bibitem [{\citenamefont {Yusupov}\ \emph {et~al.}(2010)\citenamefont
  {Yusupov}, \citenamefont {Mertelj}, \citenamefont {Kabanov}, \citenamefont
  {Brazovskii}, \citenamefont {Kusar}, \citenamefont {Chu}, \citenamefont
  {Fisher},\ and\ \citenamefont {Mihailovic}}]{yusupov2010coherent}%
  \BibitemOpen
  \bibfield  {author} {\bibinfo {author} {\bibfnamefont {R.}~\bibnamefont
  {Yusupov}}, \bibinfo {author} {\bibfnamefont {T.}~\bibnamefont {Mertelj}},
  \bibinfo {author} {\bibfnamefont {V.~V.}\ \bibnamefont {Kabanov}}, \bibinfo
  {author} {\bibfnamefont {S.}~\bibnamefont {Brazovskii}}, \bibinfo {author}
  {\bibfnamefont {P.}~\bibnamefont {Kusar}}, \bibinfo {author} {\bibfnamefont
  {J.-H.}\ \bibnamefont {Chu}}, \bibinfo {author} {\bibfnamefont {I.~R.}\
  \bibnamefont {Fisher}}, \ and\ \bibinfo {author} {\bibfnamefont
  {D.}~\bibnamefont {Mihailovic}},\ }\href@noop {} {\bibfield  {journal}
  {\bibinfo  {journal} {Nature Physics}\ }\textbf {\bibinfo {volume} {6}},\
  \bibinfo {pages} {681} (\bibinfo {year} {2010})}\BibitemShut {NoStop}%
\bibitem [{\citenamefont {Naseska}\ \emph {et~al.}(2018)\citenamefont
  {Naseska}, \citenamefont {Pogrebna}, \citenamefont {Cao}, \citenamefont {Xu},
  \citenamefont {Mihailovic},\ and\ \citenamefont
  {Mertelj}}]{naseska2018ultrafast}%
  \BibitemOpen
  \bibfield  {author} {\bibinfo {author} {\bibfnamefont {M.}~\bibnamefont
  {Naseska}}, \bibinfo {author} {\bibfnamefont {A.}~\bibnamefont {Pogrebna}},
  \bibinfo {author} {\bibfnamefont {G.}~\bibnamefont {Cao}}, \bibinfo {author}
  {\bibfnamefont {Z.}~\bibnamefont {Xu}}, \bibinfo {author} {\bibfnamefont
  {D.}~\bibnamefont {Mihailovic}}, \ and\ \bibinfo {author} {\bibfnamefont
  {T.}~\bibnamefont {Mertelj}},\ }\href@noop {} {\bibfield  {journal} {\bibinfo
   {journal} {Physical Review B}\ }\textbf {\bibinfo {volume} {98}},\ \bibinfo
  {pages} {035148} (\bibinfo {year} {2018})}\BibitemShut {NoStop}%
\bibitem [{\citenamefont {Stojchevska}\ \emph {et~al.}(2011)\citenamefont
  {Stojchevska}, \citenamefont {Kusar}, \citenamefont {Mertelj}, \citenamefont
  {Kabanov}, \citenamefont {Toda}, \citenamefont {Yao},\ and\ \citenamefont
  {Mihailovic}}]{stojchevska2011mechanisms}%
  \BibitemOpen
  \bibfield  {author} {\bibinfo {author} {\bibfnamefont {L.}~\bibnamefont
  {Stojchevska}}, \bibinfo {author} {\bibfnamefont {P.}~\bibnamefont {Kusar}},
  \bibinfo {author} {\bibfnamefont {T.}~\bibnamefont {Mertelj}}, \bibinfo
  {author} {\bibfnamefont {V.}~\bibnamefont {Kabanov}}, \bibinfo {author}
  {\bibfnamefont {Y.}~\bibnamefont {Toda}}, \bibinfo {author} {\bibfnamefont
  {X.}~\bibnamefont {Yao}}, \ and\ \bibinfo {author} {\bibfnamefont
  {D.}~\bibnamefont {Mihailovic}},\ }\href@noop {} {\bibfield  {journal}
  {\bibinfo  {journal} {Physical Review B}\ }\textbf {\bibinfo {volume} {84}},\
  \bibinfo {pages} {180507} (\bibinfo {year} {2011})}\BibitemShut {NoStop}%
\bibitem [{Note1()}]{Note1}%
  \BibitemOpen
  \bibinfo {note} {In the 3-pulse experiment the D pulse is not-chopped an
  therefore its contribution to the Pr is suppressed {[}see Eq. (\ref
  {eq:DR3}){]}.}\BibitemShut {Stop}%
\bibitem [{\citenamefont {Thomsen}\ \emph {et~al.}(1986)\citenamefont
  {Thomsen}, \citenamefont {Grahn}, \citenamefont {Maris},\ and\ \citenamefont
  {Tauc}}]{thomsen1986picosecond}%
  \BibitemOpen
  \bibfield  {author} {\bibinfo {author} {\bibfnamefont {C.}~\bibnamefont
  {Thomsen}}, \bibinfo {author} {\bibfnamefont {H.~T.}\ \bibnamefont {Grahn}},
  \bibinfo {author} {\bibfnamefont {H.~J.}\ \bibnamefont {Maris}}, \ and\
  \bibinfo {author} {\bibfnamefont {J.}~\bibnamefont {Tauc}},\ }\href {\doibase
  10.1016/0030-4018(86)90116-1} {\bibfield  {journal} {\bibinfo  {journal}
  {Optics Communications}\ }\textbf {\bibinfo {volume} {60}},\ \bibinfo {pages}
  {55} (\bibinfo {year} {1986})}\BibitemShut {NoStop}%
\bibitem [{\citenamefont {Dordevic}\ \emph {et~al.}(2001)\citenamefont
  {Dordevic}, \citenamefont {Basov}, \citenamefont {Dynes},\ and\ \citenamefont
  {Bucher}}]{dordevic2001anisotropic}%
  \BibitemOpen
  \bibfield  {author} {\bibinfo {author} {\bibfnamefont {S.~V.}\ \bibnamefont
  {Dordevic}}, \bibinfo {author} {\bibfnamefont {D.~N.}\ \bibnamefont {Basov}},
  \bibinfo {author} {\bibfnamefont {R.~C.}\ \bibnamefont {Dynes}}, \ and\
  \bibinfo {author} {\bibfnamefont {E.}~\bibnamefont {Bucher}},\ }\href
  {\doibase 10.1103/PhysRevB.64.161103} {\bibfield  {journal} {\bibinfo
  {journal} {Physical Review B}\ }\textbf {\bibinfo {volume} {64}},\ \bibinfo
  {pages} {161103} (\bibinfo {year} {2001})}\BibitemShut {NoStop}%
\bibitem [{Note2()}]{Note2}%
  \BibitemOpen
  \bibinfo {note} {The error bar is the non-linear least square fit error and
  does not take into account the index of refraction error.}\BibitemShut
  {Stop}%
\bibitem [{\citenamefont {Feldman}(1976)}]{feldman1976elastic}%
  \BibitemOpen
  \bibfield  {author} {\bibinfo {author} {\bibfnamefont {J.~L.}\ \bibnamefont
  {Feldman}},\ }\href {\doibase 10.1016/0022-3697(76)90143-8} {\bibfield
  {journal} {\bibinfo  {journal} {Journal of Physics and Chemistry of Solids}\
  }\textbf {\bibinfo {volume} {37}},\ \bibinfo {pages} {1141} (\bibinfo {year}
  {1976})}\BibitemShut {NoStop}%
\bibitem [{\citenamefont {Krasovskii}\ \emph {et~al.}(2002)\citenamefont
  {Krasovskii}, \citenamefont {Schattke}, \citenamefont {Strocov},\ and\
  \citenamefont {Claessen}}]{krasovskii2002unoccupied}%
  \BibitemOpen
  \bibfield  {author} {\bibinfo {author} {\bibfnamefont {E.~E.}\ \bibnamefont
  {Krasovskii}}, \bibinfo {author} {\bibfnamefont {W.}~\bibnamefont
  {Schattke}}, \bibinfo {author} {\bibfnamefont {V.~N.}\ \bibnamefont
  {Strocov}}, \ and\ \bibinfo {author} {\bibfnamefont {R.}~\bibnamefont
  {Claessen}},\ }\href {\doibase 10.1103/PhysRevB.66.235403} {\bibfield
  {journal} {\bibinfo  {journal} {Physical Review B}\ }\textbf {\bibinfo
  {volume} {66}},\ \bibinfo {pages} {235403} (\bibinfo {year}
  {2002})}\BibitemShut {NoStop}%
\bibitem [{Note3()}]{Note3}%
  \BibitemOpen
  \bibinfo {note} {In the case of the 3-pulse experiment we define the
  magnitude as value of the transient reflectivity at $t_{\protect \mathrm
  {PPr}}\sim 18$ ps {[}the gray bar in Fig. \ref {fig:fluence} c){]} where a
  plateau is observed in the presence of the D pulse..}\BibitemShut {Stop}%
\bibitem [{\citenamefont {Kusar}\ \emph {et~al.}(2008)\citenamefont {Kusar},
  \citenamefont {Kabanov}, \citenamefont {Demsar}, \citenamefont {Mertelj},
  \citenamefont {Sugai},\ and\ \citenamefont
  {Mihailovic}}]{kusar2008controlled}%
  \BibitemOpen
  \bibfield  {author} {\bibinfo {author} {\bibfnamefont {P.}~\bibnamefont
  {Kusar}}, \bibinfo {author} {\bibfnamefont {V.~V.}\ \bibnamefont {Kabanov}},
  \bibinfo {author} {\bibfnamefont {J.}~\bibnamefont {Demsar}}, \bibinfo
  {author} {\bibfnamefont {T.}~\bibnamefont {Mertelj}}, \bibinfo {author}
  {\bibfnamefont {S.}~\bibnamefont {Sugai}}, \ and\ \bibinfo {author}
  {\bibfnamefont {D.}~\bibnamefont {Mihailovic}},\ }\href {\doibase
  10.1103/PhysRevLett.101.227001} {\bibfield  {journal} {\bibinfo  {journal}
  {Physical Review Letters}\ }\textbf {\bibinfo {volume} {101}},\ \bibinfo
  {pages} {227001} (\bibinfo {year} {2008})}\BibitemShut {NoStop}%
\bibitem [{Note4()}]{Note4}%
  \BibitemOpen
  \bibinfo {note} {At the 1.55 eV Pr photon energy\protect \citep
  {dordevic2001anisotropic}.}\BibitemShut {Stop}%
\bibitem [{\citenamefont {Harper}\ \emph {et~al.}(1975)\citenamefont {Harper},
  \citenamefont {Geballe},\ and\ \citenamefont
  {Di~Salvo}}]{harper1975heatcapacity}%
  \BibitemOpen
  \bibfield  {author} {\bibinfo {author} {\bibfnamefont {J.~M.~E.}\
  \bibnamefont {Harper}}, \bibinfo {author} {\bibfnamefont {T.~H.}\
  \bibnamefont {Geballe}}, \ and\ \bibinfo {author} {\bibfnamefont {F.~J.}\
  \bibnamefont {Di~Salvo}},\ }\href {\doibase 10.1016/0375-9601(75)90592-7}
  {\bibfield  {journal} {\bibinfo  {journal} {Physics Letters A}\ }\textbf
  {\bibinfo {volume} {54}},\ \bibinfo {pages} {27} (\bibinfo {year}
  {1975})}\BibitemShut {NoStop}%
\bibitem [{\citenamefont {Murphy}\ \emph {et~al.}(2005)\citenamefont {Murphy},
  \citenamefont {Requardt}, \citenamefont {Stettner}, \citenamefont {Serrano},
  \citenamefont {Krisch}, \citenamefont {M{\"u}ller},\ and\ \citenamefont
  {Press}}]{murphy2005phononmodes}%
  \BibitemOpen
  \bibfield  {author} {\bibinfo {author} {\bibfnamefont {B.~M.}\ \bibnamefont
  {Murphy}}, \bibinfo {author} {\bibfnamefont {H.}~\bibnamefont {Requardt}},
  \bibinfo {author} {\bibfnamefont {J.}~\bibnamefont {Stettner}}, \bibinfo
  {author} {\bibfnamefont {J.}~\bibnamefont {Serrano}}, \bibinfo {author}
  {\bibfnamefont {M.}~\bibnamefont {Krisch}}, \bibinfo {author} {\bibfnamefont
  {M.}~\bibnamefont {M{\"u}ller}}, \ and\ \bibinfo {author} {\bibfnamefont
  {W.}~\bibnamefont {Press}},\ }\href {\doibase 10.1103/PhysRevLett.95.256104}
  {\bibfield  {journal} {\bibinfo  {journal} {Physical Review Letters}\
  }\textbf {\bibinfo {volume} {95}},\ \bibinfo {pages} {256104} (\bibinfo
  {year} {2005})}\BibitemShut {NoStop}%
\bibitem [{Note5()}]{Note5}%
  \BibitemOpen
  \bibinfo {note} {The CDW gap is concentrated on few hot spots with large
  amounts of ungapped Fermi surface.\protect \citep
  {borisenko2009twoenergy,rahn2012gapsand}}\BibitemShut {NoStop}%
\bibitem [{\citenamefont {Demsar}\ \emph {et~al.}(1999)\citenamefont {Demsar},
  \citenamefont {Biljakovi{\'c}},\ and\ \citenamefont
  {Mihailovic}}]{demsar1999singleparticle}%
  \BibitemOpen
  \bibfield  {author} {\bibinfo {author} {\bibfnamefont {J.}~\bibnamefont
  {Demsar}}, \bibinfo {author} {\bibfnamefont {K.}~\bibnamefont
  {Biljakovi{\'c}}}, \ and\ \bibinfo {author} {\bibfnamefont {D.}~\bibnamefont
  {Mihailovic}},\ }\href {\doibase 10.1103/PhysRevLett.83.800} {\bibfield
  {journal} {\bibinfo  {journal} {Physical Review Letters}\ }\textbf {\bibinfo
  {volume} {83}},\ \bibinfo {pages} {800} (\bibinfo {year} {1999})}\BibitemShut
  {NoStop}%
\bibitem [{\citenamefont {Yusupov}\ \emph {et~al.}(2008)\citenamefont
  {Yusupov}, \citenamefont {Mertelj}, \citenamefont {Chu}, \citenamefont
  {Fisher},\ and\ \citenamefont {Mihailovic}}]{yusupov2008singleparticle}%
  \BibitemOpen
  \bibfield  {author} {\bibinfo {author} {\bibfnamefont {R.}~\bibnamefont
  {Yusupov}}, \bibinfo {author} {\bibfnamefont {T.}~\bibnamefont {Mertelj}},
  \bibinfo {author} {\bibfnamefont {J.-H.}\ \bibnamefont {Chu}}, \bibinfo
  {author} {\bibfnamefont {I.}~\bibnamefont {Fisher}}, \ and\ \bibinfo {author}
  {\bibfnamefont {D.}~\bibnamefont {Mihailovic}},\ }\href@noop {} {\bibfield
  {journal} {\bibinfo  {journal} {Physical review letters}\ }\textbf {\bibinfo
  {volume} {101}},\ \bibinfo {pages} {246402} (\bibinfo {year}
  {2008})}\BibitemShut {NoStop}%
\bibitem [{\citenamefont {Stojchevska}\ \emph {et~al.}(2017)\citenamefont
  {Stojchevska}, \citenamefont {Boro{\v v}s{\v s}ak}, \citenamefont
  {{Foury-Leylekian}}, \citenamefont {Pouget}, \citenamefont {Mertelj},\ and\
  \citenamefont {Mihailovic}}]{stojchevska2017evolution}%
  \BibitemOpen
  \bibfield  {author} {\bibinfo {author} {\bibfnamefont {L.}~\bibnamefont
  {Stojchevska}}, \bibinfo {author} {\bibfnamefont {M.}~\bibnamefont {Boro{\v
  v}s{\v s}ak}}, \bibinfo {author} {\bibfnamefont {P.}~\bibnamefont
  {{Foury-Leylekian}}}, \bibinfo {author} {\bibfnamefont {J.-P.}\ \bibnamefont
  {Pouget}}, \bibinfo {author} {\bibfnamefont {T.}~\bibnamefont {Mertelj}}, \
  and\ \bibinfo {author} {\bibfnamefont {D.}~\bibnamefont {Mihailovic}},\
  }\href {\doibase 10.1103/PhysRevB.96.035429} {\bibfield  {journal} {\bibinfo
  {journal} {Physical Review B}\ }\textbf {\bibinfo {volume} {96}},\ \bibinfo
  {pages} {035429} (\bibinfo {year} {2017})}\BibitemShut {NoStop}%
\bibitem [{\citenamefont {Nasretdinova}\ \emph {et~al.}(2019)\citenamefont
  {Nasretdinova}, \citenamefont {Borov{\v s}ak}, \citenamefont {Mravlje},
  \citenamefont {{\v S}utar}, \citenamefont {Goreshnik}, \citenamefont
  {Mertelj},\ and\ \citenamefont {Mihailovic}}]{nasretdinova2019timeresolved}%
  \BibitemOpen
  \bibfield  {author} {\bibinfo {author} {\bibfnamefont {V.}~\bibnamefont
  {Nasretdinova}}, \bibinfo {author} {\bibfnamefont {M.}~\bibnamefont {Borov{\v
  s}ak}}, \bibinfo {author} {\bibfnamefont {J.}~\bibnamefont {Mravlje}},
  \bibinfo {author} {\bibfnamefont {P.}~\bibnamefont {{\v S}utar}}, \bibinfo
  {author} {\bibfnamefont {E.}~\bibnamefont {Goreshnik}}, \bibinfo {author}
  {\bibfnamefont {T.}~\bibnamefont {Mertelj}}, \ and\ \bibinfo {author}
  {\bibfnamefont {D.}~\bibnamefont {Mihailovic}},\ }\href {\doibase
  10.1103/PhysRevB.99.085101} {\bibfield  {journal} {\bibinfo  {journal}
  {Physical Review B}\ }\textbf {\bibinfo {volume} {99}},\ \bibinfo {pages}
  {085101} (\bibinfo {year} {2019})}\BibitemShut {NoStop}%
\bibitem [{\citenamefont {Kuo}\ \emph {et~al.}(2019)\citenamefont {Kuo},
  \citenamefont {Shen}, \citenamefont {Li}, \citenamefont {Quyen},
  \citenamefont {Tzeng}, \citenamefont {Luo}, \citenamefont {Wang},
  \citenamefont {Kuo},\ and\ \citenamefont {Lue}}]{kuo2019characterization}%
  \BibitemOpen
  \bibfield  {author} {\bibinfo {author} {\bibfnamefont {C.~N.}\ \bibnamefont
  {Kuo}}, \bibinfo {author} {\bibfnamefont {D.}~\bibnamefont {Shen}}, \bibinfo
  {author} {\bibfnamefont {B.~S.}\ \bibnamefont {Li}}, \bibinfo {author}
  {\bibfnamefont {N.~N.}\ \bibnamefont {Quyen}}, \bibinfo {author}
  {\bibfnamefont {W.~Y.}\ \bibnamefont {Tzeng}}, \bibinfo {author}
  {\bibfnamefont {C.~W.}\ \bibnamefont {Luo}}, \bibinfo {author} {\bibfnamefont
  {L.~M.}\ \bibnamefont {Wang}}, \bibinfo {author} {\bibfnamefont {Y.~K.}\
  \bibnamefont {Kuo}}, \ and\ \bibinfo {author} {\bibfnamefont {C.~S.}\
  \bibnamefont {Lue}},\ }\href {\doibase 10.1103/PhysRevB.99.235121} {\bibfield
   {journal} {\bibinfo  {journal} {Physical Review B}\ }\textbf {\bibinfo
  {volume} {99}},\ \bibinfo {pages} {235121} (\bibinfo {year}
  {2019})}\BibitemShut {NoStop}%
\bibitem [{\citenamefont {Mialitsin}(2011)}]{mialitsin2011fanoline}%
  \BibitemOpen
  \bibfield  {author} {\bibinfo {author} {\bibfnamefont {A.}~\bibnamefont
  {Mialitsin}},\ }\href {\doibase 10.1016/j.jpcs.2010.10.044} {\bibfield
  {journal} {\bibinfo  {journal} {Journal of Physics and Chemistry of Solids}\
  }\bibinfo {series} {Spectroscopies in {{Novel Superconductors}} 2010},\
  \textbf {\bibinfo {volume} {72}},\ \bibinfo {pages} {568} (\bibinfo {year}
  {2011})}\BibitemShut {NoStop}%
\bibitem [{\citenamefont {Demsar}\ \emph {et~al.}(2003)\citenamefont {Demsar},
  \citenamefont {Averitt}, \citenamefont {Taylor}, \citenamefont {Kabanov},
  \citenamefont {Kang}, \citenamefont {Kim}, \citenamefont {Choi},\ and\
  \citenamefont {Lee}}]{demsar2003pairbreaking}%
  \BibitemOpen
  \bibfield  {author} {\bibinfo {author} {\bibfnamefont {J.}~\bibnamefont
  {Demsar}}, \bibinfo {author} {\bibfnamefont {R.~D.}\ \bibnamefont {Averitt}},
  \bibinfo {author} {\bibfnamefont {A.~J.}\ \bibnamefont {Taylor}}, \bibinfo
  {author} {\bibfnamefont {V.~V.}\ \bibnamefont {Kabanov}}, \bibinfo {author}
  {\bibfnamefont {W.~N.}\ \bibnamefont {Kang}}, \bibinfo {author}
  {\bibfnamefont {H.~J.}\ \bibnamefont {Kim}}, \bibinfo {author} {\bibfnamefont
  {E.~M.}\ \bibnamefont {Choi}}, \ and\ \bibinfo {author} {\bibfnamefont
  {S.~I.}\ \bibnamefont {Lee}},\ }\href {\doibase
  10.1103/PhysRevLett.91.267002} {\bibfield  {journal} {\bibinfo  {journal}
  {Physical Review Letters}\ }\textbf {\bibinfo {volume} {91}},\ \bibinfo
  {pages} {267002} (\bibinfo {year} {2003})}\BibitemShut {NoStop}%
\bibitem [{\citenamefont {Beck}\ \emph {et~al.}(2011)\citenamefont {Beck},
  \citenamefont {Klammer}, \citenamefont {Lang}, \citenamefont {Leiderer},
  \citenamefont {Kabanov}, \citenamefont {Gol'tsman},\ and\ \citenamefont
  {Demsar}}]{beck2011energygap}%
  \BibitemOpen
  \bibfield  {author} {\bibinfo {author} {\bibfnamefont {M.}~\bibnamefont
  {Beck}}, \bibinfo {author} {\bibfnamefont {M.}~\bibnamefont {Klammer}},
  \bibinfo {author} {\bibfnamefont {S.}~\bibnamefont {Lang}}, \bibinfo {author}
  {\bibfnamefont {P.}~\bibnamefont {Leiderer}}, \bibinfo {author}
  {\bibfnamefont {V.~V.}\ \bibnamefont {Kabanov}}, \bibinfo {author}
  {\bibfnamefont {G.~N.}\ \bibnamefont {Gol'tsman}}, \ and\ \bibinfo {author}
  {\bibfnamefont {J.}~\bibnamefont {Demsar}},\ }\href {\doibase
  10.1103/PhysRevLett.107.177007} {\bibfield  {journal} {\bibinfo  {journal}
  {Physical Review Letters}\ }\textbf {\bibinfo {volume} {107}},\ \bibinfo
  {pages} {177007} (\bibinfo {year} {2011})}\BibitemShut {NoStop}%
\bibitem [{\citenamefont {Akiba}\ \emph {et~al.}(2023)\citenamefont {Akiba},
  \citenamefont {Toda}, \citenamefont {Tsuchiya}, \citenamefont {Oda},
  \citenamefont {Kurosawa}, \citenamefont {Mihailovic},\ and\ \citenamefont
  {Mertelj}}]{akiba2023photoinduced}%
  \BibitemOpen
  \bibfield  {author} {\bibinfo {author} {\bibfnamefont {T.}~\bibnamefont
  {Akiba}}, \bibinfo {author} {\bibfnamefont {Y.}~\bibnamefont {Toda}},
  \bibinfo {author} {\bibfnamefont {S.}~\bibnamefont {Tsuchiya}}, \bibinfo
  {author} {\bibfnamefont {M.}~\bibnamefont {Oda}}, \bibinfo {author}
  {\bibfnamefont {T.}~\bibnamefont {Kurosawa}}, \bibinfo {author}
  {\bibfnamefont {D.}~\bibnamefont {Mihailovic}}, \ and\ \bibinfo {author}
  {\bibfnamefont {T.}~\bibnamefont {Mertelj}},\ }\href@noop {} {\bibfield
  {journal} {\bibinfo  {journal} {submitted to PRB}\ } (\bibinfo {year}
  {2023})}\BibitemShut {NoStop}%
\bibitem [{\citenamefont {Rothwarf}\ and\ \citenamefont
  {Taylor}(1967)}]{rothwarf1967measurement}%
  \BibitemOpen
  \bibfield  {author} {\bibinfo {author} {\bibfnamefont {A.}~\bibnamefont
  {Rothwarf}}\ and\ \bibinfo {author} {\bibfnamefont {B.}~\bibnamefont
  {Taylor}},\ }\href@noop {} {\bibfield  {journal} {\bibinfo  {journal}
  {Physical Review Letters}\ }\textbf {\bibinfo {volume} {19}},\ \bibinfo
  {pages} {27} (\bibinfo {year} {1967})}\BibitemShut {NoStop}%
\bibitem [{\citenamefont {Kabanov}\ \emph {et~al.}(2005)\citenamefont
  {Kabanov}, \citenamefont {Demsar},\ and\ \citenamefont
  {Mihailovic}}]{kabanov2005kinetics}%
  \BibitemOpen
  \bibfield  {author} {\bibinfo {author} {\bibfnamefont {V.~V.}\ \bibnamefont
  {Kabanov}}, \bibinfo {author} {\bibfnamefont {J.}~\bibnamefont {Demsar}}, \
  and\ \bibinfo {author} {\bibfnamefont {D.}~\bibnamefont {Mihailovic}},\
  }\href@noop {} {\bibfield  {journal} {\bibinfo  {journal} {Physical review
  letters}\ }\textbf {\bibinfo {volume} {95}},\ \bibinfo {pages} {147002}
  (\bibinfo {year} {2005})}\BibitemShut {NoStop}%
\bibitem [{Note6()}]{Note6}%
  \BibitemOpen
  \bibinfo {note} {At strong excitations the saturation non linearity naturally
  leads to the drop of the rise time.}\BibitemShut {Stop}%
\bibitem [{\citenamefont {Mertelj}\ \emph {et~al.}(2009)\citenamefont
  {Mertelj}, \citenamefont {O{\v s}lak}, \citenamefont {Dolin{\v s}ek},
  \citenamefont {Fisher}, \citenamefont {Kabanov},\ and\ \citenamefont
  {Mihailovic}}]{mertelj2009finestructure}%
  \BibitemOpen
  \bibfield  {author} {\bibinfo {author} {\bibfnamefont {T.}~\bibnamefont
  {Mertelj}}, \bibinfo {author} {\bibfnamefont {A.}~\bibnamefont {O{\v s}lak}},
  \bibinfo {author} {\bibfnamefont {J.}~\bibnamefont {Dolin{\v s}ek}}, \bibinfo
  {author} {\bibfnamefont {I.~R.}\ \bibnamefont {Fisher}}, \bibinfo {author}
  {\bibfnamefont {V.~V.}\ \bibnamefont {Kabanov}}, \ and\ \bibinfo {author}
  {\bibfnamefont {D.}~\bibnamefont {Mihailovic}},\ }\href {\doibase
  10.1103/PhysRevLett.102.086405} {\bibfield  {journal} {\bibinfo  {journal}
  {Physical Review Letters}\ }\textbf {\bibinfo {volume} {102}},\ \bibinfo
  {pages} {086405} (\bibinfo {year} {2009})}\BibitemShut {NoStop}%
\bibitem [{Note7()}]{Note7}%
  \BibitemOpen
  \bibinfo {note} {Replacing the first term in brackets with $\protect \qopname
  \relax o{exp}(-t/\tau )$.}\BibitemShut {Stop}%
\bibitem [{Note8()}]{Note8}%
  \BibitemOpen
  \bibinfo {note} {With respect to the probe.}\BibitemShut {Stop}%
\bibitem [{Note9()}]{Note9}%
  \BibitemOpen
  \bibinfo {note} {The advanced diffusion model fits virtually overlap the
  simple diffusion model at 1.55 eV PPE and are not shown.}\BibitemShut {Stop}%
\bibitem [{Note10()}]{Note10}%
  \BibitemOpen
  \bibinfo {note} {In the framework of the saturation model.}\BibitemShut
  {Stop}%
\bibitem [{\citenamefont {Mertelj}\ \emph {et~al.}(2013)\citenamefont
  {Mertelj}, \citenamefont {Kusar}, \citenamefont {Kabanov}, \citenamefont
  {{Giraldo-Gallo}}, \citenamefont {Fisher},\ and\ \citenamefont
  {Mihailovic}}]{mertelj2013incoherent}%
  \BibitemOpen
  \bibfield  {author} {\bibinfo {author} {\bibfnamefont {T.}~\bibnamefont
  {Mertelj}}, \bibinfo {author} {\bibfnamefont {P.}~\bibnamefont {Kusar}},
  \bibinfo {author} {\bibfnamefont {V.~V.}\ \bibnamefont {Kabanov}}, \bibinfo
  {author} {\bibfnamefont {P.}~\bibnamefont {{Giraldo-Gallo}}}, \bibinfo
  {author} {\bibfnamefont {I.~R.}\ \bibnamefont {Fisher}}, \ and\ \bibinfo
  {author} {\bibfnamefont {D.}~\bibnamefont {Mihailovic}},\ }\href {\doibase
  10.1103/PhysRevLett.110.156401} {\bibfield  {journal} {\bibinfo  {journal}
  {Physical Review Letters}\ }\textbf {\bibinfo {volume} {110}},\ \bibinfo
  {pages} {156401} (\bibinfo {year} {2013})}\BibitemShut {NoStop}%
\bibitem [{\citenamefont {Orenstein}\ \emph {et~al.}(2023)\citenamefont
  {Orenstein}, \citenamefont {Duncan}, \citenamefont {Munoz}, \citenamefont
  {Huang}, \citenamefont {Krapivin}, \citenamefont {Nguyen}, \citenamefont
  {Teitelbaum}, \citenamefont {Singh}, \citenamefont {Mankowsky}, \citenamefont
  {Lemke}, \citenamefont {Sander}, \citenamefont {Deng}, \citenamefont
  {Arrell}, \citenamefont {Fisher}, \citenamefont {Reis},\ and\ \citenamefont
  {Trigo}}]{orenstein2023subdiffusive}%
  \BibitemOpen
  \bibfield  {author} {\bibinfo {author} {\bibfnamefont {G.}~\bibnamefont
  {Orenstein}}, \bibinfo {author} {\bibfnamefont {R.~A.}\ \bibnamefont
  {Duncan}}, \bibinfo {author} {\bibfnamefont {G.~A. d. l.~P.}\ \bibnamefont
  {Munoz}}, \bibinfo {author} {\bibfnamefont {Y.}~\bibnamefont {Huang}},
  \bibinfo {author} {\bibfnamefont {V.}~\bibnamefont {Krapivin}}, \bibinfo
  {author} {\bibfnamefont {Q.~L.}\ \bibnamefont {Nguyen}}, \bibinfo {author}
  {\bibfnamefont {S.}~\bibnamefont {Teitelbaum}}, \bibinfo {author}
  {\bibfnamefont {A.~G.}\ \bibnamefont {Singh}}, \bibinfo {author}
  {\bibfnamefont {R.}~\bibnamefont {Mankowsky}}, \bibinfo {author}
  {\bibfnamefont {H.}~\bibnamefont {Lemke}}, \bibinfo {author} {\bibfnamefont
  {M.}~\bibnamefont {Sander}}, \bibinfo {author} {\bibfnamefont
  {Y.}~\bibnamefont {Deng}}, \bibinfo {author} {\bibfnamefont {C.}~\bibnamefont
  {Arrell}}, \bibinfo {author} {\bibfnamefont {I.~R.}\ \bibnamefont {Fisher}},
  \bibinfo {author} {\bibfnamefont {D.~A.}\ \bibnamefont {Reis}}, \ and\
  \bibinfo {author} {\bibfnamefont {M.}~\bibnamefont {Trigo}},\ }\href
  {\doibase 10.48550/arXiv.2304.00168} {\enquote {\bibinfo {title}
  {Subdiffusive {{Dynamics}} of {{Topological Vortex Strings}} of a {{Charge
  Density Wave}}},}\ } (\bibinfo {year} {2023}),\ \Eprint
  {http://arxiv.org/abs/2304.00168} {arXiv:2304.00168 [cond-mat]} \BibitemShut
  {NoStop}%
\bibitem [{Note11()}]{Note11}%
  \BibitemOpen
  \bibinfo {note} {Calculating the $T$-dependent in-plane equilibrium heat
  diffusion constant from literature data\protect \citep
  {harper1975heatcapacity,beletskii1998thermal} we obtain $D_{\protect \mathrm
  {eq-ip}}\sim 1$ cm$^{2}$/s at $T_{\protect \mathrm {CDW}}$ increasing with
  decreasing $T$ to $D_{eq-\protect \mathrm {ip}}\sim 30$ cm$^{2}$/s at
  $T_{\protect \mathrm {c}}$.}\BibitemShut {Stop}%
\bibitem [{Note12()}]{Note12}%
  \BibitemOpen
  \bibinfo {note} {In the strong excitation cases, $D_{\protect \mathrm {op}}$
  is found smaller, $\sim 0.2$ and $\sim 0.01$~cm$^{2}$/s (at $T=4$~K) for the
  1.55 eV PPE and 3.1 eV PPE excitation, respectively. Here, however, one
  should take into account that the transient $T$ in the strong excitation
  cases is of the order of $T_{\protect \mathrm {CDW}}$. $D_{\protect \mathrm
  {op}}$ is therefore found $\sim 20$ times smaller (at $T\sim 26$ K) and $\sim
  40$ times smaller (at $T\sim 48$ K) than $D_{\protect \mathrm {eq-ip}}$ for
  the 1.55 eV PPE and 3.1 eV PPE excitation, respectively.}\BibitemShut {Stop}%
\bibitem [{\citenamefont {Roeske}\ \emph {et~al.}(1977)\citenamefont {Roeske},
  \citenamefont {Shanks},\ and\ \citenamefont
  {Finnemore}}]{roeske1977superconducting}%
  \BibitemOpen
  \bibfield  {author} {\bibinfo {author} {\bibfnamefont {F.}~\bibnamefont
  {Roeske}}, \bibinfo {author} {\bibfnamefont {H.~R.}\ \bibnamefont {Shanks}},
  \ and\ \bibinfo {author} {\bibfnamefont {D.~K.}\ \bibnamefont {Finnemore}},\
  }\href {\doibase 10.1103/PhysRevB.16.3929} {\bibfield  {journal} {\bibinfo
  {journal} {Physical Review B}\ }\textbf {\bibinfo {volume} {16}},\ \bibinfo
  {pages} {3929} (\bibinfo {year} {1977})}\BibitemShut {NoStop}%
\bibitem [{\citenamefont {LeBlanc}\ and\ \citenamefont
  {Nader}(2010)}]{leblanc2010resistivity}%
  \BibitemOpen
  \bibfield  {author} {\bibinfo {author} {\bibfnamefont {A.}~\bibnamefont
  {LeBlanc}}\ and\ \bibinfo {author} {\bibfnamefont {A.}~\bibnamefont
  {Nader}},\ }\href {\doibase 10.1016/j.ssc.2010.05.001} {\bibfield  {journal}
  {\bibinfo  {journal} {Solid State Communications}\ }\textbf {\bibinfo
  {volume} {150}},\ \bibinfo {pages} {1346} (\bibinfo {year}
  {2010})}\BibitemShut {NoStop}%
\bibitem [{\citenamefont {Pfalzgraf}\ and\ \citenamefont
  {Spreckels}(1987)}]{pfalzgraf1987theanisotropy}%
  \BibitemOpen
  \bibfield  {author} {\bibinfo {author} {\bibfnamefont {B.~W.}\ \bibnamefont
  {Pfalzgraf}}\ and\ \bibinfo {author} {\bibfnamefont {H.}~\bibnamefont
  {Spreckels}},\ }\href {\doibase 10.1088/0022-3719/20/27/013} {\bibfield
  {journal} {\bibinfo  {journal} {Journal of Physics C: Solid State Physics}\
  }\textbf {\bibinfo {volume} {20}},\ \bibinfo {pages} {4359} (\bibinfo {year}
  {1987})}\BibitemShut {NoStop}%
\bibitem [{\citenamefont {Schaefer}\ \emph {et~al.}(2014)\citenamefont
  {Schaefer}, \citenamefont {Kabanov},\ and\ \citenamefont
  {Demsar}}]{schaefer2014collective}%
  \BibitemOpen
  \bibfield  {author} {\bibinfo {author} {\bibfnamefont {H.}~\bibnamefont
  {Schaefer}}, \bibinfo {author} {\bibfnamefont {V.~V.}\ \bibnamefont
  {Kabanov}}, \ and\ \bibinfo {author} {\bibfnamefont {J.}~\bibnamefont
  {Demsar}},\ }\href {\doibase 10.1103/PhysRevB.89.045106} {\bibfield
  {journal} {\bibinfo  {journal} {Physical Review B}\ }\textbf {\bibinfo
  {volume} {89}},\ \bibinfo {pages} {045106} (\bibinfo {year}
  {2014})}\BibitemShut {NoStop}%
\bibitem [{Note13()}]{Note13}%
  \BibitemOpen
  \bibinfo {note} {While the observed transient reflectivity dynamics clearly
  contains some sub-picosecond-timescale components the CDW component rise time
  is significantly slower than the AM oscillation period.}\BibitemShut {Stop}%
\bibitem [{Note14()}]{Note14}%
  \BibitemOpen
  \bibinfo {note} {\protect \citet {anikin2020ultrafast} observed a $\sim
  4$-THz ($\sim 130$ cm$^{-1}$) coherent mode at 2.2 eV PPE, which softens by
  $\sim 10\%$ with increasing $T$ towards $T_{\protect \mathrm {CDW}}$. The
  data were, however, taken at rather high $F_{\protect \mathrm
  {P}}=250\protect \:\mu $J/cm$^{2}$, well above the CDW destruction threshold
  fluence, $F_{\protect \mathrm {th}}$, and the mode persists in the normal
  state so it cannot correspond to the back-folded lattice mode contributing to
  the CDW OP.}\BibitemShut {Stop}%
\bibitem [{Note15()}]{Note15}%
  \BibitemOpen
  \bibinfo {note} {At comparable volume excitation densities.}\BibitemShut
  {Stop}%
\bibitem [{Note16()}]{Note16}%
  \BibitemOpen
  \bibinfo {note} {When both diameters are much larger than the corresponding
  wavelengths.}\BibitemShut {Stop}%
\bibitem [{Note17()}]{Note17}%
  \BibitemOpen
  \bibinfo {note} {Here we implicitly assume a chosen and fixed $t_{\protect
  \mathrm {DP}}$.}\BibitemShut {Stop}%
\bibitem [{\citenamefont {Borisenko}\ \emph {et~al.}(2009)\citenamefont
  {Borisenko}, \citenamefont {Kordyuk}, \citenamefont {Zabolotnyy},
  \citenamefont {Inosov}, \citenamefont {Evtushinsky}, \citenamefont
  {B{\"u}chner}, \citenamefont {Yaresko}, \citenamefont {Varykhalov},
  \citenamefont {Follath}, \citenamefont {Eberhardt}, \citenamefont {Patthey},\
  and\ \citenamefont {Berger}}]{borisenko2009twoenergy}%
  \BibitemOpen
  \bibfield  {author} {\bibinfo {author} {\bibfnamefont {S.~V.}\ \bibnamefont
  {Borisenko}}, \bibinfo {author} {\bibfnamefont {A.~A.}\ \bibnamefont
  {Kordyuk}}, \bibinfo {author} {\bibfnamefont {V.~B.}\ \bibnamefont
  {Zabolotnyy}}, \bibinfo {author} {\bibfnamefont {D.~S.}\ \bibnamefont
  {Inosov}}, \bibinfo {author} {\bibfnamefont {D.}~\bibnamefont {Evtushinsky}},
  \bibinfo {author} {\bibfnamefont {B.}~\bibnamefont {B{\"u}chner}}, \bibinfo
  {author} {\bibfnamefont {A.~N.}\ \bibnamefont {Yaresko}}, \bibinfo {author}
  {\bibfnamefont {A.}~\bibnamefont {Varykhalov}}, \bibinfo {author}
  {\bibfnamefont {R.}~\bibnamefont {Follath}}, \bibinfo {author} {\bibfnamefont
  {W.}~\bibnamefont {Eberhardt}}, \bibinfo {author} {\bibfnamefont
  {L.}~\bibnamefont {Patthey}}, \ and\ \bibinfo {author} {\bibfnamefont
  {H.}~\bibnamefont {Berger}},\ }\href {\doibase
  10.1103/PhysRevLett.102.166402} {\bibfield  {journal} {\bibinfo  {journal}
  {Physical Review Letters}\ }\textbf {\bibinfo {volume} {102}},\ \bibinfo
  {pages} {166402} (\bibinfo {year} {2009})}\BibitemShut {NoStop}%
\bibitem [{\citenamefont {Beletskii}\ \emph {et~al.}(1998)\citenamefont
  {Beletskii}, \citenamefont {Gavrenko}, \citenamefont {Merisov}, \citenamefont
  {Obolenskii}, \citenamefont {Sologubenko}, \citenamefont {Khadjai},\ and\
  \citenamefont {Chashka}}]{beletskii1998thermal}%
  \BibitemOpen
  \bibfield  {author} {\bibinfo {author} {\bibfnamefont {V.~I.}\ \bibnamefont
  {Beletskii}}, \bibinfo {author} {\bibfnamefont {O.~A.}\ \bibnamefont
  {Gavrenko}}, \bibinfo {author} {\bibfnamefont {B.~A.}\ \bibnamefont
  {Merisov}}, \bibinfo {author} {\bibfnamefont {M.~A.}\ \bibnamefont
  {Obolenskii}}, \bibinfo {author} {\bibfnamefont {A.~V.}\ \bibnamefont
  {Sologubenko}}, \bibinfo {author} {\bibfnamefont {G.~Y.}\ \bibnamefont
  {Khadjai}}, \ and\ \bibinfo {author} {\bibfnamefont {K.~B.}\ \bibnamefont
  {Chashka}},\ }\href {\doibase 10.1063/1.593583} {\bibfield  {journal}
  {\bibinfo  {journal} {Low Temperature Physics}\ }\textbf {\bibinfo {volume}
  {24}},\ \bibinfo {pages} {273} (\bibinfo {year} {1998})}\BibitemShut
  {NoStop}%
\end{thebibliography}%

\end{document}